\documentclass[journal]{IEEEtran}

\usepackage{ifpdf}
\usepackage{cite}
\usepackage{amsmath,amssymb,amsfonts}
\usepackage{algorithmic}
\usepackage{graphicx, adjustbox}
\usepackage{array}
\usepackage{enumitem}
\usepackage{breqn}
\usepackage{ dsfont }
\usepackage{ upgreek }
\usepackage{textcomp}
\usepackage{multirow}
\usepackage{algorithm}
\usepackage{algorithmic}
\usepackage{caption}
\usepackage{pgfplots}
\usepackage{tkz-euclide}
\usepackage{subcaption}
\usepackage{tikz}{\tiny }
\usepackage{hyperref}
\usepackage{balance}
\usepackage{bigints}

\usepackage{multirow}
\newcolumntype{P}[1]{>{\centering\arraybackslash}p{#1}}

\begin{document}
	
	\title{Federated Learning for COVID-19 Detection with Generative Adversarial Networks in Edge Cloud Computing}
	
		\author{Dinh C. Nguyen,~\IEEEmembership{Member,~IEEE,}	Ming Ding,~\IEEEmembership{Senior Member,~IEEE,} Pubudu N. Pathirana,~\IEEEmembership{Senior Member,~IEEE,} \\
			Aruna Seneviratne,~\IEEEmembership{Senior Member,~IEEE}, and Albert Y. Zomaya,~\IEEEmembership{Fellow,~IEEE}

		\thanks{Dinh C. Nguyen is with the School of Engineering, Deakin University, Waurn Ponds, VIC 3216, Australia   (e-mail: cdnguyen@deakin.edu.au).}
		\thanks{Ming Ding is with Data61, CSIRO, Australia (email: ming.ding@data61.csiro.au).}
		\thanks{Pubudu N. Pathirana is with the School of Engineering, Deakin University, Waurn Ponds, VIC 3216, Australia (email: pubudu.pathirana@deakin.edu.au).}		
		\thanks{Aruna Seneviratne is with the School of Electrical Engineering and Telecommunications, University of New South Wales (UNSW), NSW, Australia (e-mail: a.seneviratne@unsw.edu.au).}
		\thanks{Albert Y. Zomaya is with the School of Computer Science, The University of Sydney, Australia (e-mail: albert.zomaya@sydney.edu.au).}
	}
	
	\markboth{Accepted at IEEE Internet of Things Journal}%
	{}

	\maketitle
	
	\begin{abstract}
	COVID-19 has spread rapidly across the globe and become a deadly pandemic. Recently, many artificial intelligence-based approaches have been used for COVID-19 detection, but they often require public data sharing with cloud datacentres and thus remain privacy concerns. This paper proposes a new federated learning scheme, called FedGAN, to generate realistic COVID-19 images for facilitating privacy-enhanced COVID-19 detection with generative adversarial networks (GANs) in edge cloud computing. Particularly, we first propose a GAN where a discriminator and a generator based on convolutional neural networks (CNNs) at each edge-based medical institution alternatively are trained to mimic the real COVID-19 data distribution. Then, we propose a new federated learning solution which allows local GANs to collaborate and exchange learned parameters with a cloud server, aiming to enrich the global GAN model for generating realistic COVID-19 images without the need for sharing actual data. {\color{black}To enhance the privacy in federated COVID-19 data analytics, we integrate a differential privacy solution at each hospital institution. Moreover, we propose a new blockchain-based FedGAN framework for secure COVID-19 data analytics, by decentralizing the FL process with a new mining solution for low running latency.}  Simulations results demonstrate the superiority of our approach for COVID-19 detection over the state-of-the-art schemes.
	\end{abstract}
	
	\begin{IEEEkeywords}
		COVID-19, federated learning, generative adversarial network,  edge cloud. 
	\end{IEEEkeywords}
	
	\IEEEpeerreviewmaketitle

\section{Introduction}
The COVID-19 pandemic has caused a devastating effect on the public health and global economy. The severity of the epidemic is enormous that the World Health Organization (WHO) has declared it as a pandemic within a month of its wide-scale expansion \cite{review4}.  Recently, artificial intelligence (AI) techniques such as machine learning (ML) \cite{li2014scaling} and deep learning (DL) \cite{phamartificial2020} have been employed to automatically diagnose and detect COVID-19 using X-ray images. For example, the projects at \cite{review2,revise1,revise2}  develop DL-based techniques such as convolutional neural networks (CNNs) \cite{dark} to identify COVID-19 cases by extracting essential features from chest X-rays. To implement DL algorithms for COVID-19-related detection, collecting large COVID-19-related X-ray images is a significant task, but it is a highly expensive and time-consuming process that requires the participation of many patients and experts. Besides, in the context of COVID-19 pandemic, the data collection becomes more challenging since medical staff may risk infections during the collection process.

Recently, generative adversarial networks (GANs) \cite{add10} have been received much attention for medical imaging applications, which allows for generating high-quality synthetic images from original data based on the interaction of two components including a generator and a discriminator via a min-max game. In this game, the generator is responsible for generating synthesized data points from random samples while the discriminator attempts to distinguish between the real samples and the one produced by the generator. The ultimate goal of training a GAN is to obtain a generator that can well capture the real data distribution to generate real-looking synthetic data. The use of GANs thus mitigates the pressure of data collection as well as improves COVID-19 data training performances \cite{7}. However, in practical scenarios, COVID-19 image datasets are distributed across multiple sites, where data in each site is too limited in quantity and diversity to train accurately a GAN for the entire data population \cite{8}. Moreover, due to the growing user privacy concerns and strict institutional regulations, the data owners, e.g., hospitals, are not willing to share their data with datacentres which thus hinders the COVID-19 analytics. This dilemma makes it hard to aggregate all distributed data in a single server to implement centralized COVID-19 data training. Therefore, developing a collaborative solution for the integration of COVID-19 data across multiple institutions is highly needed to help overcome these challenges as well as boost the COVID-19 detection performance. 

Federated learning (FL) as an emerging  distributed collaborative AI approach \cite{nguyenfederated2021}, \cite{IoT7} is particularly attractive for assisting intelligent health data analytics, including COVID-19 prediction tasks \cite{10}. This learning paradigm is enabled by coordinating multiple data sources to perform collaborative AI training with an aggregator such as a cloud server. In this way, each institution can train its data model by exchanging its model gradients without the need for sharing actual data.  This approach thus offers means to protect effectively data privacy and reduce the cost of data transmission and storage. In the COVID-19 disease scenario, we observe that data for generative learning are distributed among various medical institutions, e.g., COVID-19 X-ray images are stored in hospitals. By integrating with GANs, FL can be naturally leveraged to build a federated generative model over the distributed institutions for high-quality and privacy-enhanced COVID-19 data training. This aims to address three critical challenges, including dataset limitation, privacy protection, and constrained training performances in COVID-19 analytics. 

{\color{black}There are still several remaining issues in the current FL systems. For example, traditional FL frameworks mostly rely on a single server for the model aggregation, which possibly leads to single-point failures once the server is attacked. Moreover, since the hospitals need to use a central server for the data training, malicious attacks at the server can exploit the user information implicitly carried by  the model updates or modify the model information without authorization, which makes the FL unreliable. In this context, blockchain has emerged as a promising solution to replace the central server to coordinate the FL process, by using its decentralized networking topology \cite{IoT9}. In this context, hospitals can participate in the FL data training in a decentralized manner without the need for a central server. This solution helps avoid the risks of single-point failures and  significantly mitigates model aggregation attacks for secure and reliable data FL training.}

\subsection{Related Works}
Several studies using GAN and FL have been proposed for supporting COVID-19 detection tasks. Specifically, the authors in \cite{11}, \cite{12} developed a GAN-based approach to detect  COVID-19 X-ray images, along with using transfer learning for lung segmentation for facilitating classification. The study in \cite{13} implemented an image synthesis approach to produce high-quality and realistic COVID-19 chest tomography (CT) images for the use in DL-based semantic segmentation and classification. Moreover, the application of GAN for synthetic chest X-ray image augmentation was also investigated in \cite{14} by using a conditional GAN, which helps speed up the COVID-19 detection with enhanced classification performances. The potential of GAN in solving data limitation in the COVID-19 pandemic was demonstrated in \cite{15}, where DL-based models, e.g., Alexnet, Googlenet, and Restnet18, are employed to evaluate the synthetic data quality and perform COVID-19-related predictions. Another work in \cite{add6} employed GANs for mobility estimation in the COVID-19 pandemic under complex social contexts and constrained training datasets with multiple data sources. 

\textcolor{black}{Furthermore, the applications of FL in COVID-19 diagnosis and detection have been investigated in recent works. For example, the study in \cite{add1} proposed a novel dynamic fusion-based FL approach for diagnostic image analytics to identify COVID-19 cases. The main focus of this work is on designing a client selection mechanism which allows for deciding clients to join the FL training based on their local model performances, and developing a model fusion solution to perform FL aggregation. The research in \cite{add2} proposed an FL scheme for federated COVID region segmentation using chest computed tomography (CT) images by the collaboration of hospitals from China, Italy, and Japan. A multi-national database consisting of 1704 scans distributed among these countries is built to build a global COVID-19 detection model via a federated semi-supervised learning technique. Another work in \cite{add3} developed a federated DL framework for privacy-enhanced detection of lung abnormalities caused by COVID-19. Each institution runs a CNN model to detect lesions from COVID-19 CT images, and update the gradients to a data centre for building a generalizable, low-cost, and scalable AI model for COVID-19 disease diagnosis and management. Furthermore, the authors in \cite{add4} leveraged FL to build a COVID-19 infection screening scheme based on chest X-ray images. Several CNN-based models are employed in the FL setting, showing promising results on COVID-19 classification compared to standalone schemes without federation.  The feasibility of FL was also evaluated via real-world experiments in \cite{21} for COVID-19 X-ray image analytics  and classification.}

In terms of blockchain-based FL, the work in \cite{20} suggested an FL model with blockchain for COVID-19 CT imaging by the cooperation of multiple hospitals. The focus of this work was on developing a data normalization-based FL technique to accurately train the collaborative deep CNN model using the datasets collected from different hospitals and CT scan machines. The work in \cite{IoT1} introduced a conceptual concept of the blockchain, edge computing, and FL integration for controlling the COVID-19 pandemic. The potential of blockchain and FL was also investigated in \cite{IoT2} for health data analytics, where the benefits of the integration of these technologies were analyzed. However, no implementation and simulation results have been reported in these works \cite{IoT1}, \cite{IoT2}. Other related works in \cite{IoT3}, \cite{IoT4} proposed blockchain-based FL schemes for health Internet of Things (IoT) networks, but their roles in COVID-19 detection have not been investigated. The studies in \cite{IoT5}, \cite{IoT6} also paid attention to blockchain-based FL designs, where the proposed solutions were used to mostly address attack issues in the data communication and model aggregation. \textcolor{black}{Nonetheless, all existing works \cite{20, IoT1, IoT2, IoT3, IoT5, IoT6} have not addressed the latency issue caused by blockchain mining in the blockchain-FL systems. Moreover, the integrated design of blockchain, FL, and GANs and the investigation of this integrated model in the COVID-19 context are still missing in the above literature studies.}

\begin{table}
	\scriptsize
	\centering
	\caption{{{\color{black}The comparison of the existing works and our scheme.}}}
	{\color{black}
		\begin{tabular}{|p{2.2cm}||c|c|c|c|c|c|c|}
			\hline
			\centering \multirow{2}{*}{\textbf{Features}} &
			\multicolumn{7}{c|}{\textbf{Schemes}} \\
			&	\cite{add1}&	\cite{add2}&	\cite{20}&	\cite{IoT1}&	\cite{IoT3}&	\cite{IoT5}&	Ours \\
			\hline
			FL for COVID-19 &	\checkmark&	\checkmark&	\checkmark&	\checkmark&	&	&	\checkmark 
			\\  \hline
			Integrated FL-GAN for COVID-19&	&	&	&	&	&	&	\checkmark
			\\  \hline
			Differential privacy design &	&	&	&	&	\checkmark	&	&	\checkmark
			\\  \hline
			Decentralized FL training&	&	&	\checkmark&	\checkmark&	\checkmark&	\checkmark&	\checkmark
			\\  \hline
			Low-latency blockchain design&	&	&	&	&	&	&	\checkmark
			\\  \hline
			Integrated blockchain-FL for COVID-19 &	&	&	\checkmark&	&	&	&	\checkmark
			\\  \hline
		\end{tabular}
		\label{table:FeatureComparisons}
		\vspace{-0.1in}
	}
\end{table}

\subsection{Motivations and Our Key Contributions}
\textcolor{black}{The motivations of our work can be explained as follows. Firstly, despite these research efforts, most existing GAN algorithms \cite{11,12,13,14} for COVID-19 analytics are trained using limited and imbalanced datasets from a single institution which cannot achieve a desired COVID-19 detection accuracy. Secondly, during a pandemic, due to the increasing user privacy concerns and strict regulations, medical institutions such as hospitals are not willing to share their COVID-19 image data with a data centre for AI training, which calls for COVID-19 data training without data sharing. Thirdly, the convergence of FL and GANs is a very interesting research direction, which can achieve better COVID-19 image augmentation with privacy enhancement for better disease detection and diagnosis. However, its potential has not been explored for the COVID-19 detection domain in the open literature \cite{add1,add2,add3,add4}. Finally, how to develop a new blockchain solution for secure and low-latency federated data training is an urgent need, aiming to support efficient COVID-19 detection in the pandemic.  }

Motivated by these limitations, we here propose a novel scheme called \textit{FedGAN} for privacy-ensured and efficient COVID-19 detection by enabling a joint design of GAN and FL across medical institutions in edge cloud computing. \textcolor{black}{The key purpose of our proposed scheme is to generate high-quality synthetic image data for supporting COVID-19 detection tasks without the need for COVID-19 image data sharing.} Our proposed solution not only solves the problems of data limitation and imbalance thanks to generative learning but also enhances  COVID-19 data privacy, as well as enhances the COVID-19 detection performance due to the collaboration of multiple data sources from distributed institutions. Moreover, we propose a new blockchain-based FedGAN framework for secure COVID-19 data analytics, by decentralizing the FL process with a new mining solution for low running latency.  {\color{black}The comparison of our paper and the related works via several key features is summarized in Table~\ref{table:FeatureComparisons}.}  In a nutshell, the unique contributions of this article are highlighted as follows:
\begin{itemize}
	\item We propose a novel FedGAN scheme for COVID-19 detection, by enabling a joint design of GAN and FL across the distributed medical institutions in edge cloud computing. This model is highly effective in generating realistic COVID-19 X-ray images and thus facilitating the automatic COVID-19 detection in the current pandemic scenario with COVID-19 data scarcity at each institution. 
	\item	We propose a collaborative data augmentation scheme where a discriminator and a generator of the GAN at each edge-based institution alternatively train their model and update their trained parameters to a cloud server without disclosing the actual image samples. The proposed solution thus enables GANs to federate the training for building the global GAN model which is used to generate synthetic COVID-19 X-ray images.  {\color{black}To enhance the privacy in federated COVID-19 data analytics, we integrate a differential privacy solution at each hospital institution.} 
	\item {\color{black}We then further propose a new blockchain-based FedGAN framework for secure COVID-19 data analytics, by decentralizing the FL process over the hospital institutions. Particularly, we propose a novel mining mechanism to mitigate the mining latency caused by the blockchain adoption in the FL system that has not been addressed in the literature works. Further, we investigate the potential of blockchain-based FedGAN in COVID-19 detection scenarios with real-world datasets. }
	\item We also design an efficient CNN-based classifier which can flexibly perform COVID-19 classification in three labelled classes (COVID-19 positive, normal, and pneumonia). Finally, we implement extensive simulations to evaluate the effectiveness of our designs, showing the significant improvement of the proposed scheme in COVID-19 detection with  low running latency, compared to existing methods. 
\end{itemize}

\subsection{Paper Structure}
The remainder of this article is organized as follows. Section~\ref{Sec:Systemmodel} describes the system model and explains the standalone GAN model as the basic solution for COVID-19 image augmentation. Next, in Section~\ref{subSec:FedGAN}, we present our FedGAN model that enhances the COVID-19 image augmentation with privacy awareness and then provide its theoretical analysis. We also propose a new blockchain-based FedGAN framework for decentralized COVID-19 data analytics. We provide the simulations and evaluate the efficiency of our scheme as well as compare with other related schemes in Section~\ref{Sec:Simulate}. Finally, Section~\ref{Section:Conclusion} concludes the paper. 
\section{System Model}
\label{Sec:Systemmodel}
In this section, we describe the network model of our proposed scheme, and then analyze the baseline standalone approach in COVID-19 detection. 

\subsection{Network Model}
\begin{figure*}
	\centering
	\includegraphics [width=0.99\linewidth]{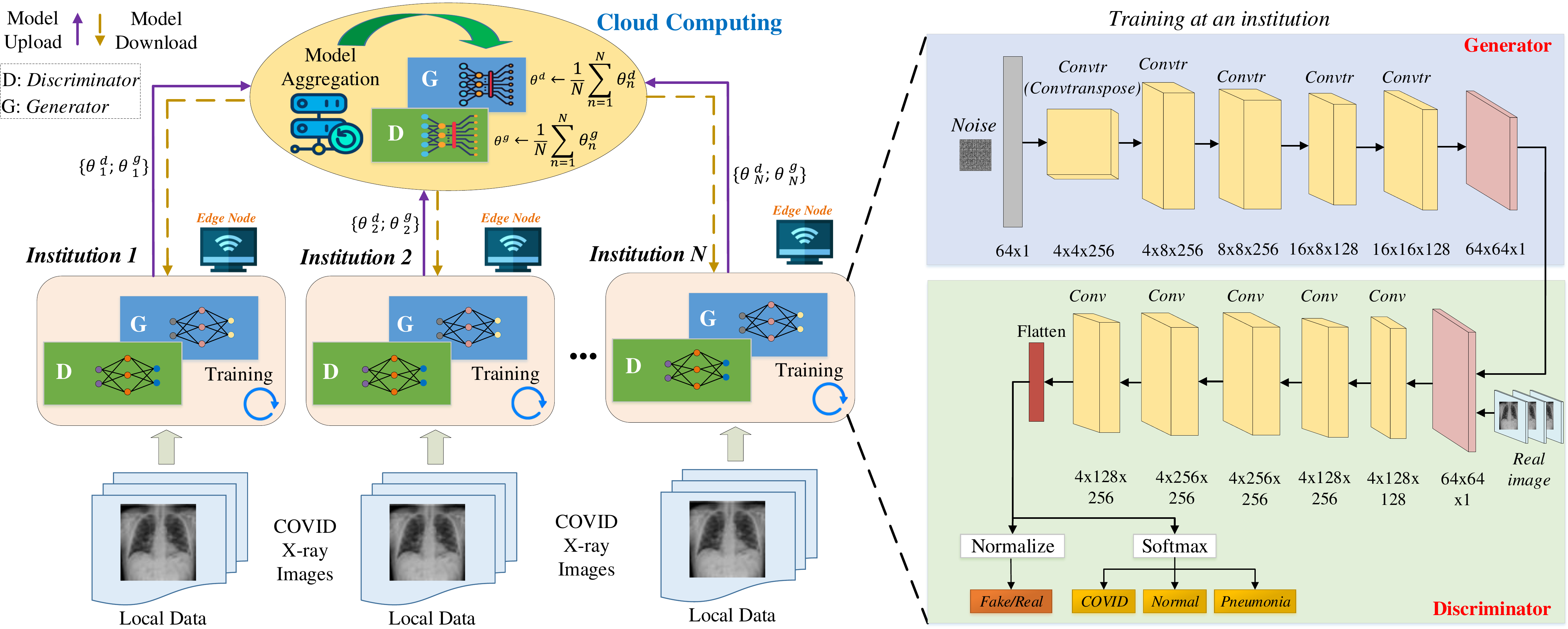}
	\caption{The proposed FedGAN architecture  for COVID-19 detection. }
	\label{Fig:Sharing}

\end{figure*}

\textcolor{black}{We consider a FedGAN model for COVID-19 detection as illustrated in Fig.~\ref{Fig:Sharing}, including a set $\mathcal{N}$ of edge nodes (ENs) located at medical institutions (e.g., hospitals) and a cloud server.  {\color{black}Note that  ENs  can be local computers or powerful IoT devices installed in hospitals for data training.}  Each EN (or institution) $n \in \mathcal{N}$ participates in the FL process using its own COVID-19 image dataset to build a global GAN with the help of a cloud server, aiming to generate high-quality synthetic COVID-19 X-ray images for improving the overall COVID-19 detection. We assume that each institution $n$ has its own dataset, denoted as $D_n$ which follows a distribution $p_n(x)$ where $x$ is the real data samples acquired from the real COVID-19 image dataset.}

We design a GAN at each EN including a generator and a discriminator by using CNNs. Particularly, each institution trains a generator to learn a generative data distribution $p^g_n$ based on its dataset $D_n$, aiming to mimic the real data distribution $p^{d}_n$ which is $p^g_n = p^{d}_n$\footnote{\textcolor{black}{At this optimal condition, the discriminator cannot distinguish the real samples from the synthetic samples generated by a well-functioning generator.}}. To do so, given a random noise $z$ from a probability distribution $p^z_n(z)$, the generator learns to generate a fake COVID-19 image data point $G_n(z,\theta_n^g)$ where $G_n$ represents the CNN with parameters $\theta_n^g$. Moreover, we design another CNN as a discriminator $D_n(x,\theta_n^d)$ at each institution which tries to classify the real COVID-19 image data point $x$ from the distribution $p_n(x)$ against the one produced from the generator. The discriminator outputs a value 1 if the input is $x$ or 0 if the input is $G_n(z,\theta_n^g)$.  Accordingly, the generator and the discriminator at each hospital interact to derive the parameters $\theta_n^g, \theta_n^d$ so that the generator produces the synthetic data distribution $p^g_n$ similar to the real data distribution $ p^{d}_n$ to fool the discriminator. Mathematically, the objective function of the GAN at each institution $n$ can be formulated via a min-max game with a value function $V_n(D,G)$ as:
\begin{multline}
\label{Equa:1}
\underset{G}{\min}~\underset{D}{\max} V_n(D,G) = \mathds{E}_{x\sim p_n^{d}} \log D_n(x) + \\ \mathds{E}_{z\sim p_n^z} \log \left(1 - D_n(G_n(z)) \right),
 \end{multline}
where $\mathds{E}$ is the expectation, $D_n(x)$ denotes the probability that $D_n$ distinguishes $x$ as real data samples, and $D_n(G_n(z))$ represents the probability that $D_n$ determines the data generated by $G_n$. In (1), the first term implies that the $D$ controls how the synthetic sample should be close to the real sample, while the second term penalizes the implausible points generated from the generator. Therefore, the discriminator $D_n$ aims to maximize the value function $V_n(D,G)$ in~\ref{Equa:1}, while the generator tries to minimize this value. 
\subsection{Standalone GAN for COVID-19 Detection}
In this subsection, we analyze the standalone GAN, a traditional approach used in \cite{11,12,13} for COVID-19 detection. In this case, every institution $n$ only trains the GAN using its own dataset without the federation. 
\\
\textbf{Proposition 1:} \textit{For a given generator $G_n$, the optimal discriminator is}
\begin{equation}
\label{Equa:D_n_optimal}
D^*_n = \frac{p^{d}_n}{p^{d}_n + p^g_n}.
\end{equation}
\\
\textit{Proof.} For a given generator $G_n$, we can derive the probability distribution function for the generator $p^g_n$. Based on \cite{NIPS2014}, we can express the value function $V_n$ in~\ref{Equa:1} as below
\begin{multline}
\label{Equa:Vn}
V_n(D_n,G_n) = \int_{x} p^{d}_n (x) \log D_n(x) dx + \\ \int_{z} p^z_n (z) \log (1-D_n(G_n(z))) dz \\ = \int_{x} \left[ p^{d}_n (x) \log D_n(x) + p^g_n (x) \log (1-D_n(x)) \right] dx.
\end{multline}
We know that the function $a \log(y) + b \log(1-y)$ achieves its minimum value at $\frac{a}{a+b}$, $\forall (a,b) \in (0,1]$, which thus leads to the result in~\ref{Equa:D_n_optimal}. Next, we derive the optimum of the standalone GAN.
\\
\textbf{Theorem 1:} \textit{The optimal value of a standalone GAN for COVID-19 detection is }
\begin{equation}
\label{Equa:Value_Standalone}
V^*_n(D_n,G_n) = - \log 4 + 2*JSD(p^{d}_n \parallel p^g_n),
\end{equation}
where $JSD(p^{d}_n \parallel p^g_n)$ is the Jensen-Shannon divergence between distributions $p^{d}_n$ and $p^g_n$ \cite{NIPS2014}. 
\\
\textit{Proof.} From \ref{Equa:D_n_optimal} and \ref{Equa:Vn}, we have
\begin{multline} 
\scriptstyle
V^*_n(D,G) = \bigintssss_{x} \left[ p^{d}_n(x) \log D^*_n(x) + p^g_n (x) \log (1-D^*_n(x)) \right] dx \\ \scriptstyle = 
\bigintssss_{x} \left[ p^{d}_n(x) \log \left(\frac{p^{d}_n(x)}{p^{d}_n(x) + p^g_n(x)}\right) + p^g_n(x) \log \left(\frac{p^g_n}{p^{d}_n + p^g_n(x)}\right) \right] dx \\  \scriptstyle = 
\bigintssss_{x} \left[ p^{d}_n(x) \log \left(\frac{p^{d}_n(x)}{2\frac{p^{d}_n(x) + p^g_n(x)}{2}}\right) + p^g_n (x) \log \left(\frac{p^g_n (x)}{2\frac{p^{d}_n(x) + p^g_n (x)}{2}}\right) \right] dx \\ \scriptstyle = 
\bigintssss_{x} p^{d}_n(x) log \frac{1}{2}dx + \bigintssss_{x} p^g_n(x) log \frac{1}{2}dx \\ \scriptstyle 
+ 
\bigintssss_{x}  p^{d}_n(x) \log \left(\frac{p^{d}_n(x)}{\frac{p^{d}_n(x) + p^g_n(x)}{2}}\right) dx + \bigintssss_{x} p^g_n (x) \log \left(\frac{p^g_n}{\frac{p^{d}_n(x) + p^g_n(x)}{2}}\right) dx \\ \scriptstyle = 
- \log 4 + KL\left(p^{d}_n \parallel \frac{p^{d}_n(x) + p^g_n(x)}{2}\right) + KL\left(p^g_n \parallel \frac{p^{d}_n(x) + p^g_n(x)}{2} \right),
\end{multline}
where $KL$ is the Kullback-Leibler divergence. \textcolor{black}{Based on \cite{NIPS2014}, we know that the Jensen-Shannon divergence between probability distributions $A(x)$ and $B(x)$ is $JSD(A \parallel B)$ which is a symmetrized and smoothed version of the all important divergence measure of Kullback-Leibler divergence $KL~(~A~\parallel~B)$ which is defined as: $KL~(~A~\parallel~C) +  KL(B\parallel C) = 2* JSD(A\parallel B)$, where $C = \frac{1}{2}(A+B)$. Accordingly, from~\ref{Equa:Value_Standalone} we derive $V^*_n(D,G) = - \log 4 + 2*JSD(p^{d}_n||p^g_n)$, completing the proof. }

\textcolor{black}{Since $JSD(p^{d}_n||p^g_n)$ is always non-negative \cite{NIPS2014}, the global minimum of $V^*_n(D,G)$ in the standalone GAN-based COVID-19 detection is $-\log 4$.} In the following, we will present the proposed FedGAN and prove theoretically its advantages in training the GAN for efficient COVID-19 detection.

\section{Proposed FedGAN for COVID-19 Detection}
\label{subSec:FedGAN}
\subsection{Theoretical Analysis of FedGAN}
In FedGAN, the value function can be defined as a multi-agent game of discriminators and generators of all institutions $n \in \mathcal{N}$. The generators cooperatively learn to generate fake COVID-19 X-ray images in order to fool all discriminators of institutions, whereas the discriminators attempt to differentiate the real data from fake images generated by generators. Then, the value function of the FedGAN can be defined as
\begin{multline}
\label{Equa:federatedValue}
V^{fed}(D,G) = \sum_{n=1}^{N} V_n(D_n,G_n) = \\ \sum_{n=1}^{N} \left[\mathds{E}_{x\sim p_n^{d}} \log D_n(x) + \mathds{E}_{z\sim p_n^z} \log \left(1 - D_n(G_n(z)) \right) \right].
\end{multline}
Moreover, for any given generators $G_n$, the optimal discriminator $D_n^*$ for the FedGAN is also similar to the standalone GAN which is given by~\ref{Equa:D_n_optimal}. Accordingly, the optimal value of the FedGAN for COVID-19 detection can be calculated as follows:
\begin{multline} 
\label{Equa_FedValueFunction}
\scriptstyle
V^{fed*}_n(D,G) =  \sum_{n=1}^{N} \bigintssss_{x} \left[ p^{d}_n(x) \log D^*_n(x) + p^g_n (x) \log (1-D^*_n(x)) \right] dx \\ \scriptstyle= 
\sum_{n=1}^{N} \bigintssss_{x} \left[ p^{d}_n(x) \log \left(\frac{p^{d}_n(x)}{p^{d}_n(x) + p^g_n(x)}\right) + p^g_n(x) \log \left(\frac{p^g_n}{p^{d}_n + p^g_n(x)}\right) \right] dx \\ \scriptstyle= 
\sum_{n=1}^{N} \bigintssss_{x} \left[ p^{d}_n(x) \log \left(\frac{p^{d}_n(x)}{2\frac{p^{d}_n(x) + p^g_n(x)}{2}}\right) + p^g_n (x) \log \left(\frac{p^g_n (x)}{2\frac{p^{d}_n(x) + p^g_n (x)}{2}}\right) \right] dx \\ \scriptstyle =  \sum_{n=1}^{N} \left[ - \log 4 + 2*JSD(p^{d}_n||p^g_n)\right] = -n\log 4 + 2*\sum_{n=1}^{N} JSD(p^{d}_n||p^g_n).
\end{multline}

\textcolor{black}{In the multi-agent game, each generator at each institution $n$ is trained to learn perfectly the contribution of the actual data $ p^{d}_n$ where the minimum of the value function $V^{fed}(D,G)$ can be achieved at $ p^{d}_n = p^g_n$ for $n \in \mathcal{N}$. Thus, the solution of \ref{Equa_FedValueFunction} yields $ \sum_{n=1}^{N}JSD(p^{g}_n||p^g_n) = 0$. As a result, the optimal value function for FedGAN can achieve as $V^{fed*}_n(D,G) = -n\log 4$. }

\textcolor{black}{Based on the above analysis, we can see that the federated approach can achieve a better global minimum of the GAN value function,  compared to the standalone approach, due to the federation of multiple institutions which enables learning the data distribution over the entire population. In order words, the proposed FedGAN approach can learn better the COVID-19 image distribution to produce better synthetic image data which facilitates the detection tasks. We will provide extensive simulations to verify the advantage of the FedGAN. }
\subsection{Training of FedGAN for COVID-19 Detection}
We denote the global training iteration horizon as $T$ and index time by $t$. Each institution $n \in \mathcal{N}$ joins the FedGAN training with the cloud server, by updating the parameters of the discriminator and the generator $\theta^d_{n,t}$ and $\theta^g_{n,t}$ in each global round and exchange them with the cloud server for aggregation. We assume that the COVID-19 image data in this work is independent and identically distributed (iid) across the institutions, while the non-iid data case will be considered in future works. For every global epoch $t$, each institution $n$ collaboratively trains its discriminator $D_n$ and generator $G_n$. Specifically, the generator $G_n$ produces $k$ minibatchs of fake samples from the noise probability distribution $p^z_n(z)$ as $\{ z^{(1)},z^{(2)},...,z^{(k)}\}$. Also, the discriminator $D_n$ samples $k$ minibatchs of real data from the actual image distribution $p^{d}_n(x)$ as $\{ x^{(1)},x^{(2)},...,x^{(k)}\}$. Then, each institution $n$  updates simultaneously the discriminator $D_n$  by ascending its stochastic gradient:
\begin{equation}
\label{Equa:Dis_update}
\bigtriangledown_{\theta^d} \frac{1}{k} \left[ \sum_{j=1}^{k} \log D\left(x^{(j)}\right) + \log \left(1 - D\left(G \left(z^{(j)}\right) \right)\right)   \right], 
\end{equation}
and updates the generator $G_n$ by descending its stochastic gradient:
\begin{equation}
\label{Equa:Gen_update}
\bigtriangledown_{\theta^g} \frac{1}{k}  \sum_{j=1}^{k} \log \left(1 - D\left(G \left(z^{(j)}\right) \right)\right),
\end{equation}
to update its own weights $\theta_n^d, \theta_n^g$. \textcolor{black}{These stochastic gradient calculations also characterize the approximation of the value function defined in~\ref{Equa:federatedValue}.}

\textcolor{black}{After the training, the institutions upload their updates $\theta_n^d, \theta_n^g$ to the cloud server for model aggregation. In this work, we adopt the popular model averaging approach \cite{10} \textcolor{black}{to aggregate the local model parameters:} $\theta^d \leftarrow  \frac{1}{N} \sum_{1}^{N} \theta_n^d$, $\theta^g \leftarrow  \frac{1}{N} \sum_{1}^{N} \theta_n^g$. Then, the cloud server broadcasts the new global updates $\theta^d, \theta^g$ to all institutions for the next round of GAN learning. The FedGAN process is iterated until the global loss function converges with a desired accuracy.} The training procedure of the proposed FedGAN is summarized in Algorithm~\ref{Al:algorithm}.
\begin{algorithm}
	
	\caption{Training procedure of the proposed FedGAN}
	\begin{algorithmic}[1]
		\label{Al:algorithm}
		\STATE \textbf{Cloud server executes:}
		\STATE Initialize global training period $T$, local training epoch $L$, local weights $\theta_n^d, \theta_n^g, \forall n \in \mathcal{N}$, learning rate $\sigma$
		\FOR {each global round $t = 1,2,...,T$}
		\FOR {each institution $n \in \mathcal{N}$}
		\STATE $\theta^d_{n,t+1}, \theta^g_{n,t+1} \leftarrow \textbf{LocalUpdate}(n, \theta_t^d, \theta_t^g )  $
		\ENDFOR
		\STATE $\theta^d_{t+1} \leftarrow  \frac{1}{N} \sum_{1}^{N} \theta^d_{n,t+1}$
		\STATE $\theta^g_{t+1} \leftarrow  \frac{1}{N} \sum_{1}^{N} \theta^g_{n,t+1}$
		\ENDFOR
		\STATE $\textbf{LocalUpdate}(n, \theta^d, \theta^g ) $\textbf{:}  \textit{// Run at each institution $n$}
		\FOR {each local epoch $i = 1,2,...,L$}
		\STATE Sample $k$ minibatchs of noise samples $\{z^{(1)},z^{(2)},...,z^{(k)}\}$ from distribution $p^z_n(z)$
		\STATE Sample $k$ minibatchs of real data  $\{ x^{(1)},x^{(2)},...,x^{(k)}\}$ from actual COVID-19 image distribution $p^{d}_n(x)$
		\STATE Update the discriminator via \ref{Equa:Dis_update}
		\STATE Update the generator via \ref{Equa:Gen_update}
		\STATE Update the weights $\theta^d$ and $\theta^g$
		\begin{equation}
		\theta^d \leftarrow \theta^d - \sigma \bigtriangledown_{\theta^d} (\theta^d,\theta^g); \theta^g \leftarrow \theta^g - \sigma \bigtriangledown_{\theta^g} (\theta^d,\theta^g)
		\end{equation}
		\ENDFOR 
		\STATE Return $\theta^d, \theta^g$ to the cloud server
	\end{algorithmic}
\end{algorithm}

In FedGAN, the computational complexity mostly comes from the computation at each institution since the cloud server only implements the parameter aggregation and  does not result in much computational costs. Thus, we here focus on analyzing the computational complexity at each institution. In every global training round, each institution $n$ collaboratively trains its discriminator and generator to compute its $\theta_n^d$ and $\theta_n^g$, respectively. To do that, the generator $G_n$ produces $k$ minibatchs of fake samples of batch size $b^g$, and the discriminator $D_n$ samples $k$ minibatchs of real data of batch size $b^d$. Accordingly, in the generator, the batch generation requires $ kb^gG_{fp}$ floating point operations, where $G_{fp}$ is the number of floating operations to generate a fake data sample of the generator, and a memory of $kb^gG_{neu}$, where $ G_{neu}$ is the number of neurons of the generator. Therefore, the computational complexity of the generator can be determined as $kb^gG_{fp}  + kb^gG_{neu} = \mathcal{O} \left(kb^g(G_{fp}  + G_{neu})\right)$. Similarly, the computational complexity of the discriminator can be specified  as $ \mathcal{O} \left(kb^d(D_{fp}  + D_{neu})\right)$, where $D_{fp}$ and $ D_{neu}$ are the number of floating operations to generate a real data sample and the number of neurons of the generator. To sum up, the computational complexity at an institution in the FedGAN framework after $T$ global training round is $ \mathcal{O} \left(Tk \left[\left(b^g(G_{fp}  + G_{neu})\right) + \left(b^d(D_{fp}  + D_{neu})\right) \right]\right)$.

{\color{black}\subsection{Differential Privacy for FedGAN}

To further enhance privacy for FedGAN training, we integrate an $\epsilon$-differential privacy solution at each hospital site, where $\epsilon$ is the distinguishable bound of all outputs on two adjacent datasets $D, D'$ in a database. A randomized function $\mathcal{A}$ is $\epsilon$-differential privacy if
\begin{equation}
Pr[\mathcal{A}(D) \in \mathcal{S}] \leq e^{\epsilon}Pr[\mathcal{A}(D') \in \mathcal{S}], 
\end{equation}
where $\mathcal{S} \in range(\mathcal{A})$. \textcolor{black}{To guarantee $\epsilon$-differential privacy, we here apply a gradient perturbation technique with differentially-private stochastic gradient descent (DP-SGD) \cite{IoT7}, where Gaussian noises are added to the gradient during the training.} Accordingly, we can determine  the gradient descent update at training round $t$ as
\begin{equation}
\label{Equa:DP_update}
\theta_{t+1} = \theta_t - \sigma\left( \nabla L(\theta_t) + \zeta\right), 
\end{equation}
where $L$ is the loss function of GAN and $\zeta$ is the noise guaranteeing differential privacy.}

\section{{\color{black}Blockchain-based Federated Learning for COVID-19 Detection}}
{\color{black}Although FedGAN can support privacy-enhanced COVID-19 data analytics, how to address the security issues in terms of information leakage caused by the malicious cloud third party and single-point failures is a critical challenge. More specifically, the traditional FL framework basically relies on a single cloud server to perform FL aggregation. However, the cloud party may illegally exploit the information uploaded from hospitals without the consent of healthcare users, which potentially leads to the leakage of sensitive health information. Adversaries also deploy attacks to steal or modify the FL updates during the aggregation process at the cloud sever. Moreover, the centralized configuration in traditional FL systems is also vulnerable to the risks of single-point failures if the cloud server is attacked which would disrupt the entire FL process \cite{IoT9}. 

The roles of blockchain for secure COVID-19 data analytics have been reviewed in our recent work \cite{nguyen2021}, where blockchain can enable decentralized data learning over distributed institutions without the need for a central server. Therefore, we here  present a novel decentralized FedGAN framework by integrating a blockchain-based solution, as illustrated in Fig.~\ref{Fig:FL_Blockchain-COVID}. Instead of relying on a centralized cloud server to coordinate the FL aggregation, we here replace it with a blockchain that is able to decentralize the FL aggregation for security enhancement \cite{9}. \textcolor{black}{ Moreover, the use of blockchain is able to enhance the scalability of FL implementation in practical healthcare networks. This is enabled by its decentralization feature which allows for interconnecting distributed edge nodes and hospitals to train image datasets in a peer-to-peer manner. }
\subsection{Working Procedure of Blockchain-based FedGAN }
The working procedure of our blockchain-based FedGAN framework is explained in the following steps:
\begin{figure*}
	\centering
	\includegraphics [width=0.85\linewidth]{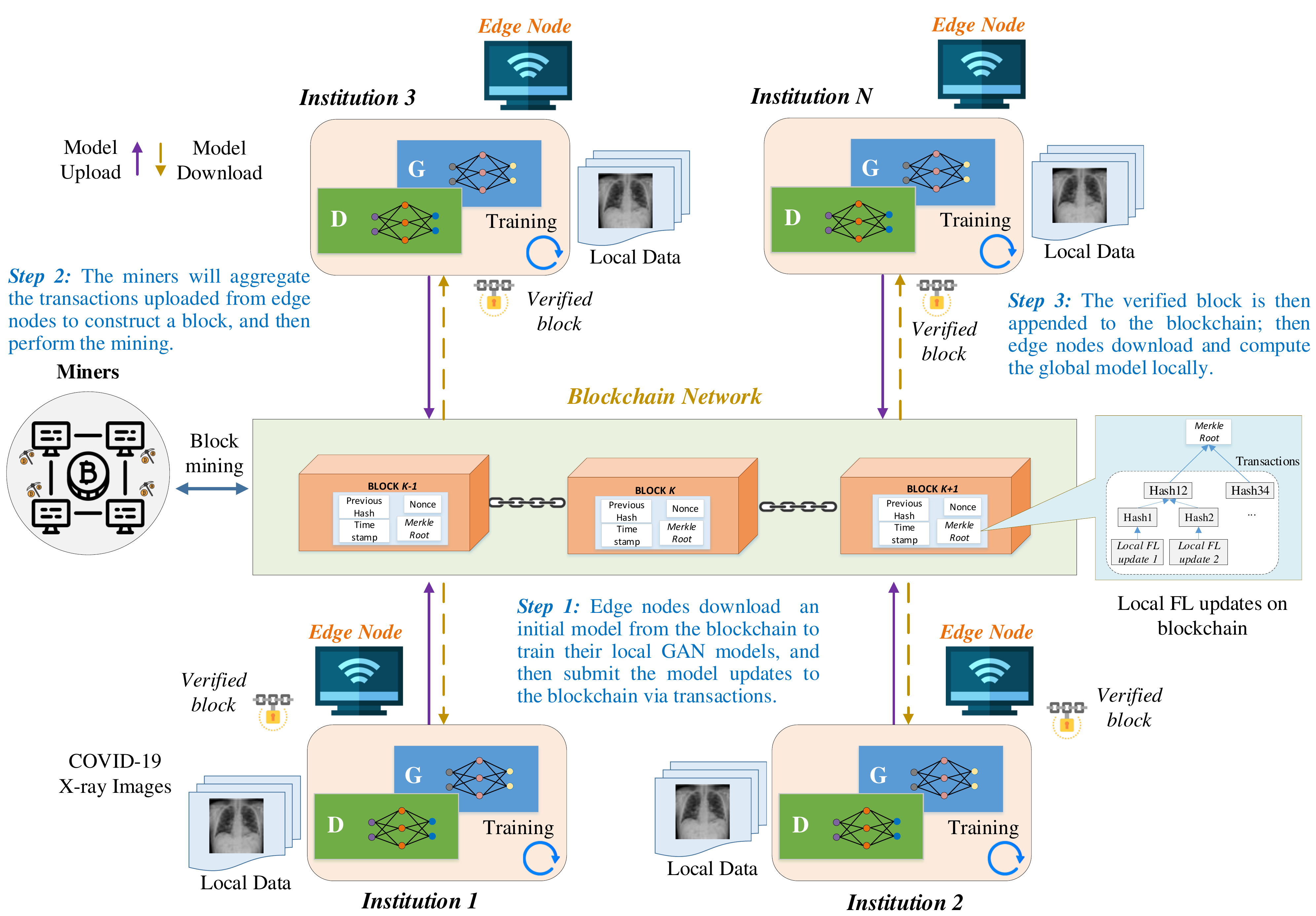}
	\caption{{\color{black}The proposed blockchain-based federated learning for COVID-19 detection.} }
	\label{Fig:FL_Blockchain-COVID}
\end{figure*}

\begin{enumerate}
	\item The EN that is willing to join the FL process downloads an initial model from the blockchain. Note that each EN also needs to set up a wallet account that contains a public key for identification and a private key for transaction signature.
	\item Each EN performs the training of the GAN model (i.e., a generator and a discriminator) to compute the gradients using its own local COVID-19 X-ray image dataset. \textcolor{black}{In the case of using differential privacy, an amount of $\epsilon$-differential privacy noises is added to the gradient during the training.} After local training, each EN submits its model updates to the blockchain by creating a transaction.
	\item  The miners will aggregate the transactions uploaded from ENs to construct a block after a certain period of time. Then, the miners perform the mining to verify the block using a consensus mechanism. 
	\item  After the mining, if all miners achieve an agreement on the verified block, this block is then appended to the blockchain. Now each EN can download the block that contains all FL updates of other ENs to compute the global model. In this regard, the global model is constructed  locally instead of in the central cloud server like in the traditional FL architecture. The GAN training is iterated until the desired accuracy performance is achieved.
\end{enumerate}
Based on the working procedure, we can see that the total running latency costs of blockchain-based FedGAN training mostly come from the FedGAN training latency and blockchain mining latency. In this particular work, we focus on addressing the blockchain mining latency, by proposing a novel block consensus mechanism as presented in the following. }

{\color{black}\subsection{Proposed Consensus Mechanism for Blockchain-based FedGAN }
\label{{Section:PoR}}
In the blockchain-based FedGAN system, when the number of transactions (e.g., local GAN updates) to the blockchain increases, the consensus workload to validate and append them into the blockchain also increases significantly. Although consensus mechanisms such as Delegated Proof-of-Stake (DPoS) have been applied to replace computationally expensive consensus schemes like Proof-of-Work, these solutions still have high latency costs. Indeed, in current consensus algorithms, e.g., DPoS \cite{consensus1}, each miner must contact at least more than half of the total nodes in the miner group, which consequently increases latency and alleviates the scalability of the blockchain system. Moreover, each miner node must implement a repeated verification process across the miner network, which results in unnecessary consensus latency. A possible solution is to reduce the number of miner nodes to mitigate the consensus latency, but it potentially compromises the security of blockchain because of the higher probability of adding compromised transactions from malicious nodes \cite{consensus1}. To solve these mining issues, here we propose a new lightweight consensus mechanism called Proof of Reputation (PoR) for our blockchain-FedGAN system. Compared to the DPoS scheme, here we make a significant  improvement in the miner selection based on a reputation score evaluation approach. Moreover, instead of using a repeated verification among miner nodes, we implement a lightweight block verification solution that allows each miner to only verify once with another node during the consensus process, which would significantly reduce the verification latency. There are two main parts to our PoR consensus, including miner node selection and block verification.
\subsubsection{Miner Node Selection}
In this phase, the ENs first calculate the reputation score of miners and then select the miner nodes to implement the mining process.

\textit{{- Reputation Calculation:}} In our blockchain-based FedGAN system, in addition to COVID-19 data training, ENs also participate in the delegate selection process to vote the mining candidates for performing blockchain consensus. In this regard, each EN votes its preferred miners with the most reputation. Here, the reputation of a miner is measured by its computing capability to mine the block. That is, a miner that allocates more computational resources to the mining tasks will have a higher reputation score to obtain a higher priority for mining the block. To this end, we define a reputation function to determine the score for each miner as follows:
\begin{equation}
\Psi_m = e^{1-\frac{T_m^{PoR}}{\tau}} -1.
\end{equation}
Here, $T_m^{PoR}$ is the mining latency of the miner $m$  where $m$ is the miner index in the mining group with $m \in \{1,...,M\}$, and $\tau$ denotes the mining latency threshold. \textcolor{black}{This equation implies that the miner that has lower mining latency will achieve a higher reputation score to obtain a higher priority for mining the block.}

\textit{- Miner Selection:} Based on the calculated reputation score, each EN votes for miner candidates based on their reputation ranking. The top miners in the mining group with highest reputation scores are selected to become actual miners (\textit{here, we call them as edge miners (EMs)}) to perform the mining. Besides, similar to the traditional DPoS framework \cite{4}, each of the EMs also acts as a block manager which is responsible for performing block generation, broadcasting blocks to other miners for verification, and block aggregation after being verified, during its time slot of the consensus process.  
\subsubsection{Lightweight Block Verification}

The block manager first generates an unverified block that contains several health transactions aggregated in a certain time period, and then transmits this block to all EMs for verification. Different from the traditional DPoS scheme which relies on a repeated verification process among miners, here we implement a lightweight PoR-based verification solution that allows each miner only needs to verify once with another node during the consensus process, which  significantly reduces the verification latency.  The block manager first divides the block $B$ with the whole transaction into $K$ transaction parts $Tr_k$ ($k = (1,...,K)$) that will be assigned to each EM member $EM_m$ within the miner group. Each miner $EM_m$ will be also assigned a unique random number $R_m$. Subsequently, a $EM_m$ selects any miner $s$ ($s!=n$) for verification on its assigned transaction part $Tr_k$. If a majority of EMs (at least 51\%) returns positive outcomes, the block manager approves the verified block $B$ and appends it into the chain. 


\subsubsection{Latency of Block Verification}
\label{Subsection:PoRLatency}
In this sub-section, we calculate the verification latency incurred by the mining. Here, we assume that each EM receives the same transaction part $Tr_k$.  Each EM is willing to contribute their resource $C = \{c_1,..., c_m\}$ (in CPU cycles/s) to execute the verification of the transaction part $k$. For each EM $n$, the CPU resource occupied to verify the transaction $k$ is $\Phi_m$. We also denote $Tr_k^{re}$ as the size of verified transaction result $Tr_k$. 

Conceptually, the block verification process in our proposed PoR mechanism at an EM experiences four steps: (1) unverified block transmission from the block manager to the EMs, (2) local block verification at the EM, (3) broadcasting of the verification result among two EMs,  and (4) transmission of verification result feedback from the EMs to the manager. The delay caused by the execution of these steps at each miner $n$ can be calculated as:
\begin{equation}
\label{EquationPoR}
T_m^{PoR} = \frac{Tr_k}{r_m^d} + \frac{\Phi_m}{c_m} +\xi Tr_k|L^2|  + \frac{Tr_k^{re}}{r_m^u}, k \in [1,...,K],
\end{equation}
where $r_m^u$ and $r_m^d$ represents the transmission rates of miner-manager uplink and downlink, respectively. Here, the transmission latency of the transaction part $Tr_k$ is $ \frac{Tr_k}{r_m^d}$, and the latency for local verification is $ \frac{\Phi_k}{c_m}$. The latency for transaction broadcasting among two miners is specified as $\xi Tr_k|L^2|$, where $\xi$ is a pre-defined parameter of transaction broadcasting among two miners and can be determined via historical verification records \cite{consensus2}. The last component is verification feedback time, shown as $ \frac{Tr_k^{re}}{r_m^u} $. 

Meanwhile, in the DPoS model \cite{4}, each miner needs to perform repeated verification on the whole block $B$, where its verification latency is computed as \cite{consensus2}:
\begin{equation}
\label{EquationDPoS}
T_m^{DPoS} = \frac{B}{r_m^d} + \frac{\Phi_m^B}{c^B_m} +\xi B|L^N|  + \frac{B^{re}}{r_m^u},
\end{equation}
where $\Phi_m^B$ represents the CPU resource needed  for executing the block $B$ with respect to the total budget $c_m^B$. Moreover, $B^{re}$ is the size of the verified outcome for block $B$. $|L^N|$ implies that all miners $n$ join the repeated verification on the block.
By comparison of equations~\ref{EquationPoR}~and~\ref{EquationDPoS}, it can be seen that the proposed PoR scheme consumes less time in the verification process, compared to the traditional DPoS scheme, for the same block size and number of miners.  The benefits of our proposed PoR mechanism will be verified in the following section. }

\section{Experiments and Performance Evaluations}
\label{Sec:Simulate}
\subsection{Experimental Settings}
\begin{figure}[t!]
	\centering
	\begin{subfigure}[t]{0.24\textwidth}
		\centering
		\includegraphics[width=0.99\linewidth]{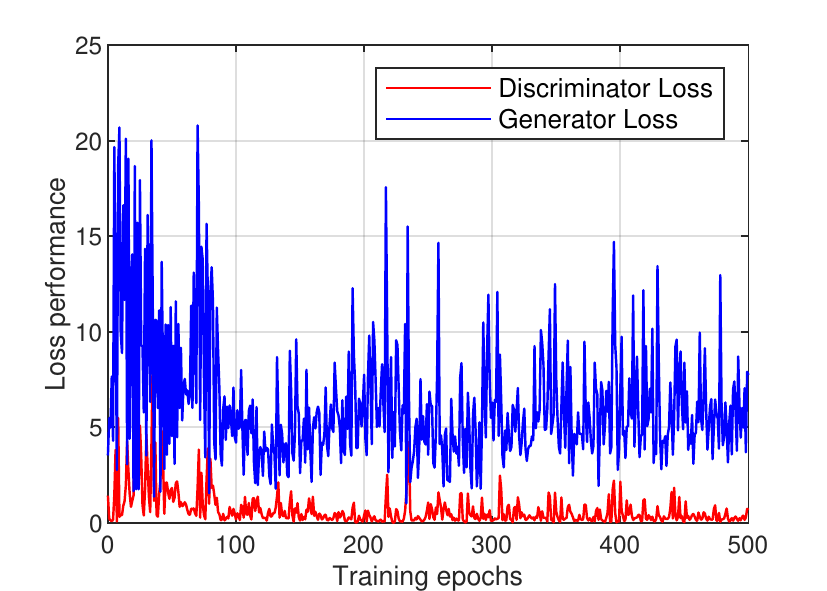} 
		\caption{FedGAN loss on training DarkCOVID dataset. }
	\end{subfigure}%
	~
	\begin{subfigure}[t]{0.24\textwidth}
		\centering
		\includegraphics[width=0.99\linewidth]{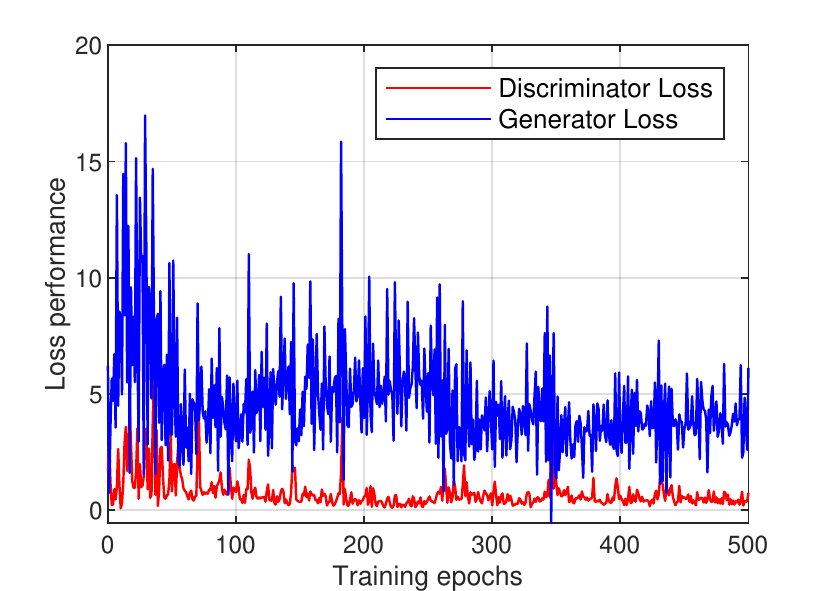} 
		\caption{FedGAN loss on training ChestCOVID dataset.}
	\end{subfigure}
	\caption{Evaluation of training loss.}
	\label{Convergence_loss_overall}
	\vspace{-0.1in}
\end{figure}

\begin{figure}[t!]
	\centering
	\begin{subfigure}[t]{0.24\textwidth}
		\centering
		\includegraphics[width=0.99\linewidth]{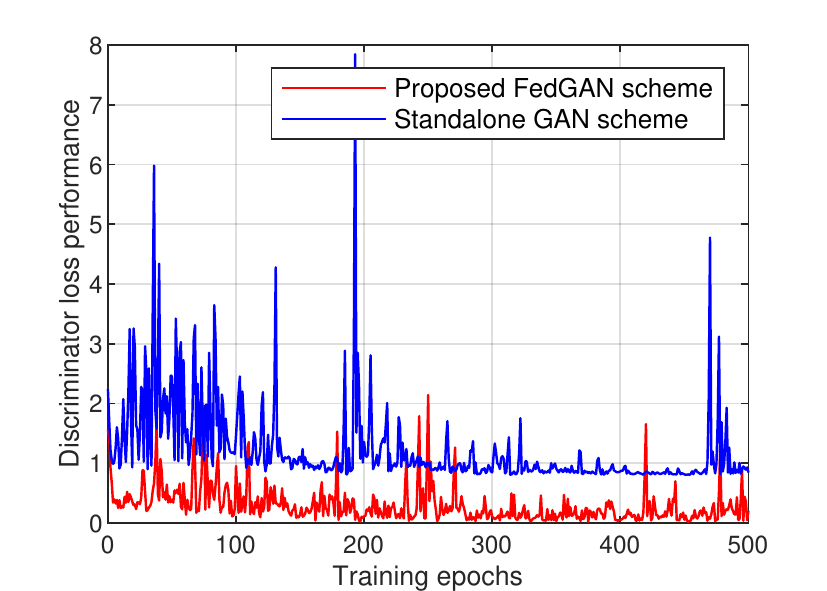} 
		\caption{Discriminator losses on training DarkCOVID dataset.}
	\end{subfigure}%
	~
	\begin{subfigure}[t]{0.24\textwidth}
		\centering
		\includegraphics[width=0.99\linewidth]{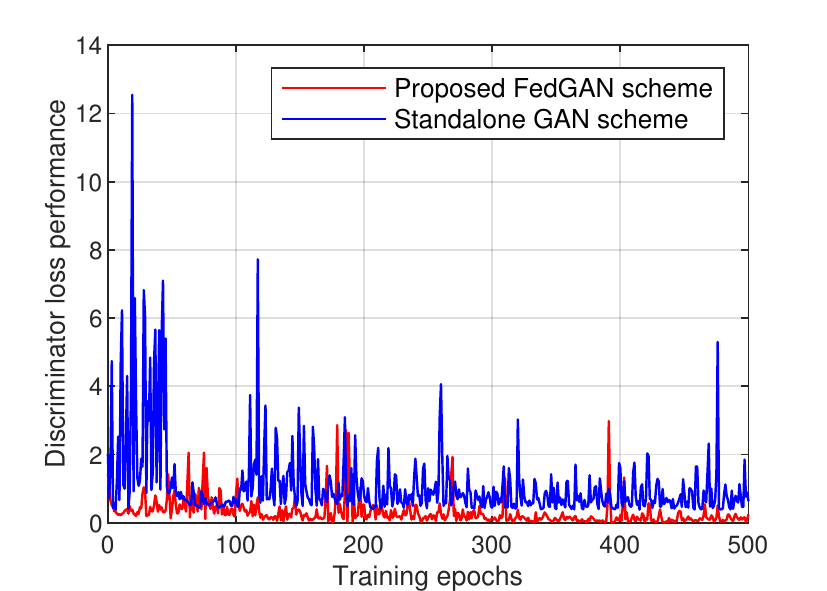} 
		\caption{Discriminator losses on training ChestCOVID dataset.}
	\end{subfigure}
	\caption{Comparison of discriminator losses.}
	\label{Convergence_loss}
	\vspace{-0.1in}
\end{figure}

\begin{figure}
	\centering
	\begin{subfigure}[b]{0.45\textwidth}
		\includegraphics[width=0.99\linewidth]{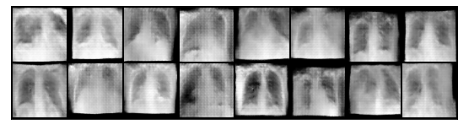}
		\caption{Synthetic COVID-19 X-ray images generated from the standalone GAN scheme.}
		\label{fig:Ng2}
	\end{subfigure}
	
	\begin{subfigure}[b]{0.45\textwidth}
		\includegraphics[width=0.99\linewidth]{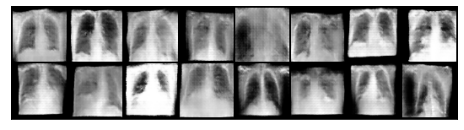}
		\caption{Synthetic COVID-19 X-ray images generated from the proposed FedGAN scheme.}
		\label{fig:Ng1} 
	\end{subfigure}
	\caption[s]{Generation of synthetic COVID-19 X-ray images. The quality of synthetic images generated from our FedGAN scheme is much better than those from the standalone scheme. }
	\label{Fig:FigureGenerated}
\end{figure}
We use two popular COVID-19 X-ray datasets for simulations, including a DarkCOVID dataset \cite{dark} with total 620 X-ray images and a ChestCOVID dataset \cite{chest}  with total 950 X-ray images for three classes (COVID-19, normal (no pneumonia), and pneumonia (with no COVID-19 infection)) which have been collected from different regions which makes them suitable for our FedGAN setting. For each dataset, we divide into the training and testing set with a 80:20 ratio.  

In the FedGAN system, we set up five institutions where each of them has a discriminator and a generator based on CNNs, as shown in Fig.~\ref{Fig:Sharing}. Each discriminator $D$ takes COVID-19 X-ray images in the form of 64x64x1 size where the mini-batch is set to 32. The discriminator is configured with five hidden layers, each having 128 dimensions along with LeakyRELU and dropout. Furthermore, each generator $G$ takes 64-dimensional noise samples from a standard Gaussian distribution. Every generator has five hidden layers, where the first three layers have 256 dimensions and the last two layers have 128 dimensions. These parameters are selected based on preliminary experimental results. The FedGAN is trained for 500 global rounds where each local GAN is trained for 20 epochs. 

Moreover, we design a CNN-based classifier for COVID-19 detection with three classes: COVID-19 positive, normal, and pneumonia. The CNN architecture consists of \textcolor{black}{an input layer with the shape of 32x32x3} and three hidden convolutional layers with kernel 3x3, ReLU activation functions and max pooling. Here, the first hidden layer has  32 dimensions, the second layer has 128 dimensions with Relu as the activation in each layer. The final layer has three dimensions, with SGD as the optimizer, a softmax function as the activation to output prediction results over three classes. Moreover, batch normalization and dropout (0.5) are added to avoid overfitting on the training set. The CNN classifier is trained with the learning rate of 0.001. These hyperparameters are selected via multiple training trials for reliable classification results. All simulations were implemented in Pytorch on a desktop server with an Intel Core i7 4.7GHz CPU and 128 GB memory with Nvidia Pascal Titan X and CUDA 8.0. 

\subsection{Performance Evaluations on FedGAN}

\begin{table}
	\centering
	\caption{Generation of synthetic COVID-19 X-ray image data using the FedGAN model.  }
	\label{Table:generaticCOVID}
	\begin{tabular}{|P{2.3cm}|P{1.5cm}|P{1.5cm}|P{1.5cm}|}
		\hline
		\textbf{Dataset}& 	
		\textbf{Classes} &	
		\textbf{Original data}&	
		\textbf{Synthetic data}
		\\
		\hline
		\multirow{4}{*}{DarkCOVID dataset} & 
		COVID-19&	150&	500
		\\ \cline{2-4}&
		Normal  &	232	&500
		\\ \cline{2-4}&
		Pneumonia	&238	&500
		\\ \cline{2-4}&
		\textbf{Sum}&	\textbf{620}&	\textbf{1500}
		\\ \cline{2-4}
		\hline
		
		\multirow{4}{*}{ChestCOVID dataset} & 
		COVID-19 &	223	&800
		\\ \cline{2-4}&
		Normal  &	421	&800
		\\ \cline{2-4}&
		Pneumonia&	306	&800
		\\ \cline{2-4}&
		\textbf{Sum}	&\textbf{950}	&\textbf{2400}
		\\ \cline{2-4}
		\hline
	\end{tabular}
	\vspace{-0.1in}
\end{table}

\begin{table}[ht]
	\centering
	\caption{Accuracies of CNN classifier on mixed actual data and synthetic data on DarkCOVID dataset. }
	\label{table:DarkCOVID_Accuracy}		
	\setlength{\tabcolsep}{5pt}
	\begin{tabular}{|c|c|c|c|c|c|}
		\hline
		\textbf{Training size}& \centering
		\textbf{$\alpha = 0$}&	   
		\textbf{$\alpha = 1$}&
		\textbf{$\alpha = 2$}&
		\textbf{$\boldsymbol{\alpha = 3}$}&
		\textbf{$\alpha = 4$}
		\\
		\hline
		500&	0.487	&0.550 &0.842&	\textbf{0.911}&	0.901
		\\
		\hline
		1000&0522		&	0.614 &0.850&	\textbf{0.925}&	0.934
		\\
		\hline
		1500&	0.663&	0.749&	0.871	&\textbf{0.931}	&0.927
		\\
		\hline
		2000&	0.784&	0.799&	0.872&	\textbf{0.967}&	0.955
		\\
		\hline
	\end{tabular}
	\vspace{-0.1in}
\end{table}
\begin{table}[ht]
	\centering
	\caption{Accuracies of CNN classifier on mixed actual data and synthetic data on ChestCOVID dataset. }
	\label{table:ChestCOVID_Accuracy}		
	\setlength{\tabcolsep}{5pt}
	\begin{tabular}{|c|c|c|c|c|c|c|}
		\hline
		\textbf{Training size}& \centering
		\textbf{$\beta = 0$}&	   
		\textbf{$\beta = 1$}&
		\textbf{$\beta = 2$}&
		\textbf{$\beta = 3$}&
		\textbf{$\boldsymbol{\beta = 4}$}&
		\textbf{$\beta = 5$}
		\\
		\hline
		500&	0.561&	0.617&	0.715&	0.890&	\textbf{0.931}&	0.900
		\\
		\hline
		1000&0.547&	0.621&	0.745&	0.846&	\textbf{0.945}&	0.918
		\\
		\hline
		1500&0.57&	0.732&	0.870&	0.871&	\textbf{0.950}&	0.925
		\\
		\hline
		2000&0.61&	0.808&	0.904&	0.932&	\textbf{0.962}&	0.953
		\\
		\hline
		2500&0.690&	0.893&	0.945&	0.945&	\textbf{0.973}&	0.932
		\\
		\hline
	\end{tabular}
	\vspace{-0.1in}
\end{table}
We investigate the performance of our proposed FedGAN scheme and compare it with the state-of-the-art schemes, including: the standalone scheme \cite{revise1}, the standalone scheme with GAN \cite{14}, the FL scheme without GAN \cite{add4}, \textcolor{black}{and the centralized scheme (all datasets are transmitted to the cloud for classification).} \textcolor{black}{All considered schemes use a CNN-based classifier for evaluating the COVID-19 detection. For reliable evaluation, the reported results are averaged from five runs of numerical simulations.}


\begin{table*}
	\scriptsize
	\centering
	\caption{\textcolor{black}{Comparison of performance results for COVID-19 detection on ChestCOVID dataset.}}
	\begin{tabular}{|p{1.2cm}||ccc|ccc|ccc|ccc|}
		\hline
		\centering \multirow{2}{*}{\textbf{Classes}} &
		\multicolumn{3}{c|}{\textbf{Standalone scheme without GAN}} & \multicolumn{3}{c|}{\textbf{Standalone scheme with GAN}}  &
		\multicolumn{3}{c|}{\textbf{FL scheme without GAN}} & \multicolumn{3}{c|}{\textbf{Proposed FedGAN scheme}} \\ 		&Precision&	Sensitivity &	F1-score 
		&Precision&	Sensitivity &	F1-score
		&Precision&	Sensitivity &	F1-score
		&Precision&	Sensitivity &	F1-score
		\\ \hline
		COVID-19 &0.871&	0.848&	0.907	&0.909	&0.893&	0.903&	0.975&	0.979&	0.988&	0.993&	0.978&	0.991
		\\  \hline
		Normal &0.867&	0.997	&0.877	&0.876&	0.959&	0.884&	0.946&	0.999&	0.948&	0.964&	1	&0.969
		\\  \hline
		Pneumonia &	0.847&	0.550&	0.696&	0.857&	0.542&	0.689&	0.941	&0.808	&0.891&	0.966&	0.876&	0.932
		\\ \hline
	\end{tabular}
	\label{table:Performance_ChestCOVIDdataset}
	\vspace{-0.1in}
\end{table*}
\begin{figure*}[t!]
	\centering
	\begin{subfigure}[t]{0.23\textwidth}
		\centering
		\includegraphics[width=0.99\linewidth]{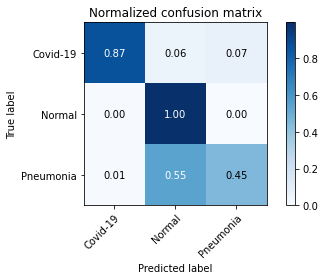} 
		\caption{Standalone scheme without GAN. }
	\end{subfigure}%
	~
	\begin{subfigure}[t]{0.23\textwidth}
		\centering
		\includegraphics[width=0.99\linewidth]{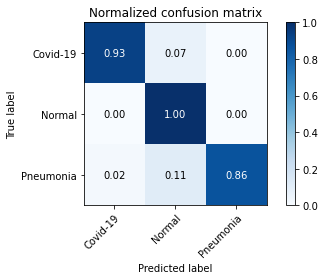} 
		\caption{Standalone scheme with GAN.}
	\end{subfigure}
	~
	\begin{subfigure}[t]{0.23\textwidth}
		\centering
		\includegraphics[width=0.99\linewidth]{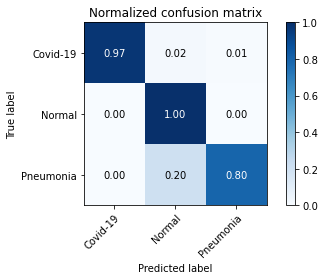} 
		\caption{FL scheme without GAN.}
	\end{subfigure}
	~
	\begin{subfigure}[t]{0.23\textwidth}
		\centering
		\includegraphics[width=0.99\linewidth]{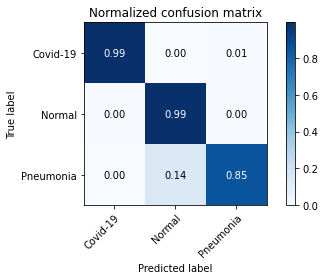} 
		\caption{Proposed FedGAN scheme.}
	\end{subfigure}
	\caption{Confusion matrixes of different schemes for COVID-19 detection on DarkCOVID dataset. }
	\label{Fig:Performance_DarkCOVIDdataset}
	\vspace{-0.1in}
\end{figure*}
\begin{figure*}[t!]
	\centering
	\begin{subfigure}[t]{0.23\textwidth}
		\centering
		\includegraphics[width=0.99\linewidth]{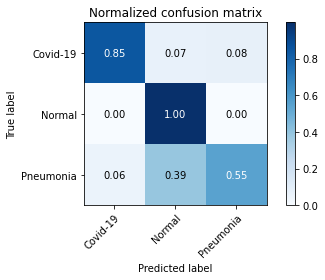} 
		\caption{Standalone scheme without GAN. }
	\end{subfigure}%
	~
	\begin{subfigure}[t]{0.23\textwidth}
		\centering
		\includegraphics[width=0.99\linewidth]{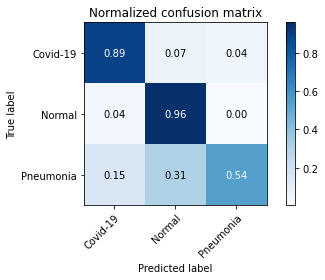} 
		\caption{Standalone scheme with GAN.}
	\end{subfigure}
	~
	\begin{subfigure}[t]{0.23\textwidth}
		\centering
		\includegraphics[width=0.99\linewidth]{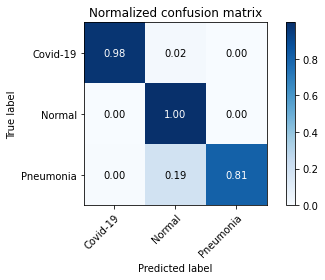} 
		\caption{FL scheme without GAN.}
	\end{subfigure}
	~
	\begin{subfigure}[t]{0.23\textwidth}
		\centering
		\includegraphics[width=0.99\linewidth]{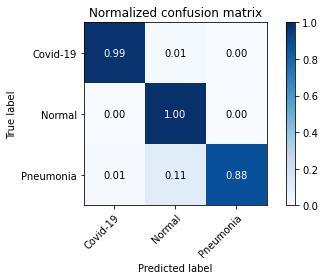} 
		\caption{Proposed FedGAN scheme.}
	\end{subfigure}
	\caption{Confusion matrixes of different schemes for COVID-19 detection on ChestCOVID dataset. }
	\label{Fig:Performance_ChestCOVIDdataset}
	\vspace{-0.13in}
\end{figure*}

\subsubsection{Evaluation of FedGAN Training}
We first evaluate the training loss of the FedGAN model, including the discriminator and generator losses for training on DarkCOVID and ChestCOVID datasets during 500 epochs, as shown in Fig.~\ref{Convergence_loss_overall}. Notably, the discriminator loss achieves a stable convergence after 100 epochs at both datasets, which show that the FedGAN model can synthesize COVID-19 image data and its quality is improved over the training time. That is, the FedGAN model can learn the features of real COVID-19 X-ray images and generate high-quality synthetic COVID-19 X-ray images in a fashion that the discriminator cannot differentiate them from the actual ones. 

In Fig.~\ref {Convergence_loss}, we compare the discriminator loss of the proposed FedGAN and the standalone scheme with GAN \cite{14}. It can be seen that the standalone scheme cannot achieve a good result in both cases due to the lack of access to the full dataset.  By contrast, our FedGAN scheme can achieve better minimum loss thanks to its ability to learn data over the entire distribution of all institutions. These simulated results are also aligned with our theoretical analysis in Section~\ref{subSec:FedGAN}. Therefore, our model can produce better COVID-19 X-ray images, as illustrated in Fig.~\ref{Fig:FigureGenerated}. The details of data generation for both datasets with  synthetic COVID-19 X-ray image numbers are presented in TABLE~\ref{Table:generaticCOVID}. Here, we generate 1500 synthetic DarkCOVID images and 2400 synthetic ChestCOVID images, with equal image volume in each class which thus addresses dataset limitation and imbalance. We will use these synthetic data associated with real data in the following simulations for COVID-19 detection.

\subsubsection{Evaluation of COVID-19 Detection Peformance with FedGAN} 
\begin{figure*}[t!]
	\centering
	\begin{subfigure}[t]{0.48\textwidth}
		\centering
		\includegraphics[width=0.99\linewidth]{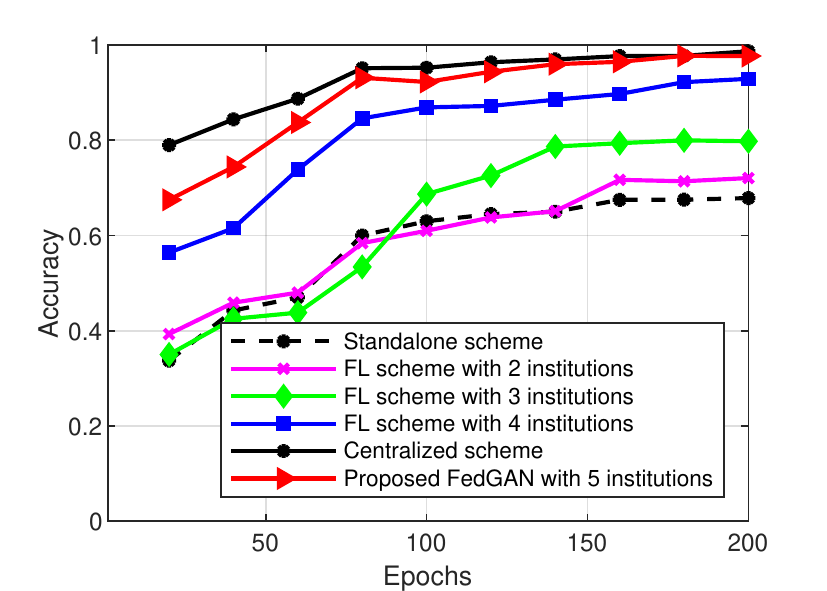} 
		\caption{Accuracy performance for COVID-19 detection on testing DarkCOVID dataset. }
	\end{subfigure}%
	~
	\begin{subfigure}[t]{0.48\textwidth}
		\centering
		\includegraphics[width=0.99\linewidth]{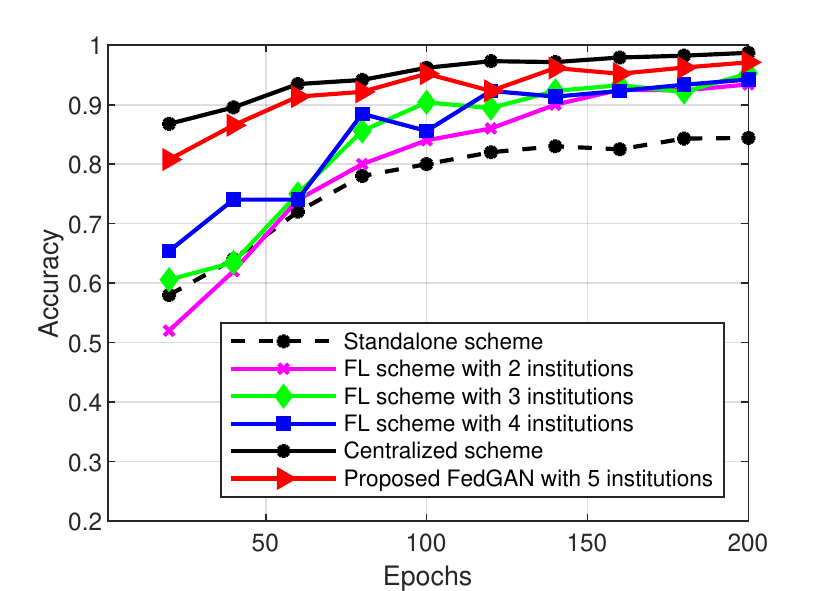} 
		\caption{Accuracy performance for COVID-19 detection on testing ChestCOVID dataset. }
	\end{subfigure}
	\caption{\textcolor{black}{Comparison of accuracy performances with different FL schemes.}}
	\label{Convergence_Accuracy}
	\vspace{-0.1in}
\end{figure*}

\begin{figure*}[t!]
	\centering
	\begin{subfigure}[t]{0.48\textwidth}
		\centering
		\includegraphics[width=0.99\linewidth]{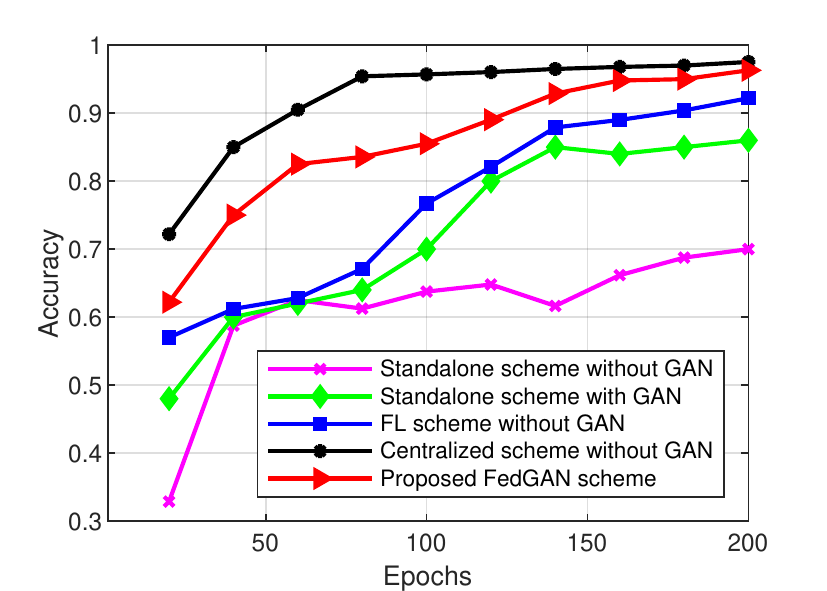} 
		\caption{Accuracy performance for COVID-19 detection on testing DarkCOVID dataset. }
	\end{subfigure}%
	~
	\begin{subfigure}[t]{0.48\textwidth}
		\centering
		\includegraphics[width=0.99\linewidth]{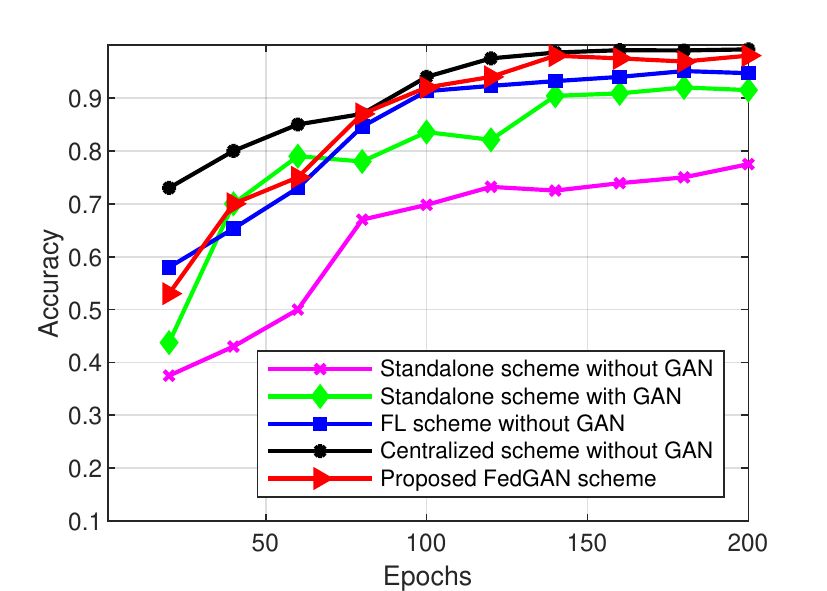} 
		\caption{Accuracy performance for COVID-19 detection on testing ChestCOVID dataset. }
	\end{subfigure}
	\caption{\textcolor{black}{Comparison of accuracy performances with different COVID-19 detection schemes.}}
	\label{Convergence_Accuracy_schemes}
	\vspace{-0.1in}
\end{figure*}
To implement COVID-19 detection, an important step is to determine how much synthetic data should be used for the best detection rate. To do so, we mix actual data and synthetic data according to different ratios as the number of synthetic data against the number of actual data, which are denoted as $\alpha$ and $\beta$ on DarkCOVID and ChestCOVID datasets, respectively. $\alpha =0$ and $\beta =0$ imply that only actual data is used for training. From TABLE~\ref{table:DarkCOVID_Accuracy}, when the mixing ratio $\alpha$ increases, the accuracies of CNN classifier generally increase on the training DarkCOVID subsets, which also shows the benefit of data augmentation offered by GANs in COVID-19 classification. Notably,  $\alpha =3$ yields the best accuracy in most training DarkCOVID sets, but increasing further synthetic data can degrade the accuracy performance due to the overfitting issue. Similarly, we also investigate on the ChestCOVID dataset in TABLE~\ref{table:ChestCOVID_Accuracy}, showing that the highest accuracy performance is achieved with $\beta = 4$. Therefore, we will use these mixing ratios for the remaining simulations.

Next, we evaluate the performance of COVID-19 detection via common quality metrics including precision, sensitivity, and F1-score. As illustrated in TABLE~\ref{table:Performance_ChestCOVIDdataset}, our proposed FedGAN scheme outperforms  other frameworks at three metrics. For instance, for the COVID-19 class in the ChestCOVID dataset, our scheme can increase the precision and F1 score up to 0.993 and 0.991, compared to lower results at other schemes.   The advantages of our FedGAN in COVID-19 detection are also confirmed via confusion matrixes in Fig.~\ref{Fig:Performance_DarkCOVIDdataset} and Fig.~\ref{Fig:Performance_ChestCOVIDdataset}. 

Furthermore, we evaluate different schemes in terms of FCN scores. They are used to measure the quality of the generated images on an input segmentation map that can be implemented by feeding generated X-ray images into the fully-convolutional semantic segmentation network (FCN). We use three standard segmentation metrics following CycleGAN \cite{long2015fully} to evaluate FCN scores, including the per-pixel accuracy, the per-class accuracy, and the mean class Intersection-Over-Union (IOU). As indicated in Table~\ref{table:DarkCOVID_FCN}, our FedGAN scheme outperforms the other approaches for both datasets. For example, in the training of DarkCOVID dataset, our scheme can improve the per-pixel accuracy by 29\%, increase the per-class accuracy by 18\% and the mean IOU by 12\%, compared to the FL scheme without GAN. The improvements on FCN scores of our scheme over existing schemes are also shown on the training of ChestCOVID dataset in Table~\ref{table:ChestCOVID_FCN}, showing the better stability of image-label translation of our proposed approach.

\begin{table}[ht]
	\centering
	\caption{\textcolor{black}{Comparison of FCN scores on DarkCOVID dataset. }}
	\label{table:DarkCOVID_FCN}		
	\setlength{\tabcolsep}{5pt}
	\begin{tabular}{|p{2.5cm}|c|c|c|c|}
		\hline
		\textbf{Schemes}&    Per-pixel acc. &	Per-class acc. &	Mean IOU
		\\
		\hline
		Standalone scheme without GAN&	0.32&	0.24&	0.28
		\\
		\hline
		Standalone scheme with GAN&	0.47&	0.41&	0.35
		\\
		\hline
		FL scheme without GAN&		0.63&	0.52&	0.49
		\\
		\hline
		Proposed FedGAN scheme&	\textbf{0.82}&	\textbf{0.65}&	\textbf{0.56}
		\\
		\hline
	\end{tabular}
	\vspace{-0.1in}
\end{table}

\begin{table}[ht]
	\centering
	\caption{\textcolor{black}{Comparison of FCN scores on ChestCOVID dataset. }}
	\label{table:ChestCOVID_FCN}		
	\setlength{\tabcolsep}{5pt}
	\begin{tabular}{|p{2.5cm}|c|c|c|c|}
		\hline
		\textbf{Schemes}&    Per-pixel acc. &	Per-class acc. &	Mean IOU
		\\
		\hline
		Standalone scheme without GAN&	0.45&	0.40&	0.34
		\\
		\hline
	Standalone scheme with GAN&	0.59&	0.58&	0.39
		\\
		\hline
		FL scheme without GAN&	0.75&	0.63&	0.43
		\\
		\hline
	Proposed FedGAN scheme&	\textbf{0.87}&	\textbf{0.74}&	\textbf{0.59}
		\\
		\hline
	\end{tabular}
	\vspace{-0.1in}
\end{table}

We then investigate the detection performance in terms of accuracy for different FL schemes and our FedGAN scheme on testing datasets. The standalone scheme is used as the baseline which theoretically has the lowest accuracy rates due to the lack of federation. As shown in Fig.~\ref{Convergence_Accuracy}, the more participating institutions in data training, the higher accuracy achieved for both datasets. This can be explained by the enhanced image feature learning efficiency thanks to the use of diverse data sources for improving the generalizability of CNN model when the number of institutions increase. However, our FedGAN scheme achieves the best accuracy among all FL approaches and \textcolor{black}{is close to the centralized scheme with the full dataset}. For example, for testing DarkCOVID dataset in Fig.~\ref{Convergence_Accuracy}(a), when the epoch is 200, the accuracy of our scheme stands at 0.992 which is 8\%, 19\%, 25\%, and 28\% higher than those of the FL schemes with 2,3,4 institutions and the standalone scheme, respectively. \textcolor{black}{
It also achieves a competitive performance level with the ideal centralized scheme (0.995).} The accuracy of our scheme is also the highest on the testing ChestCOVID dataset among baselines in Fig.~\ref{Convergence_Accuracy}(b), \textcolor{black}{achieving a value of 0.985 at 200 epochs and closeness to the centralized scheme.} 

\textcolor{black}{Moreover, we compare the accuracy performance of our scheme with other COVID-19 detection schemes, as indicated in Fig.~\ref{Convergence_Accuracy_schemes}. Our FedGAN scheme can significantly improve the accuracy rate in both datasets due to its GAN and federated learning combination. That is, our scheme yields the best accuracy of 0.963 after 200 running epochs for testing DarkCOVID dataset in Fig.~\ref{Convergence_Accuracy_schemes}(a), compared to the FL scheme without GAN (0.922), the standalone scheme with GAN (0.856), and the standalone scheme without GAN (0.705). \textcolor{black}{Its performance is also nearly close to the centralized scheme (0.969).}}  A similar observation is also obtained on testing ChestCOVID dataset in  Fig.~\ref{Convergence_Accuracy_schemes}(b), with the notable accuracy performance (0.975) achieved by our proposed FedGAN scheme. These simulation results demonstrate the high confidence and effectiveness of our design in detecting COVID-19. 

{\color{black}\subsubsection{Evaluation of Differential Privacy-enabled FedGAN Performance} 
\begin{figure}[t!]
	\centering
	\begin{subfigure}[t]{0.24\textwidth}
		\centering
		\includegraphics[width=0.99\linewidth]{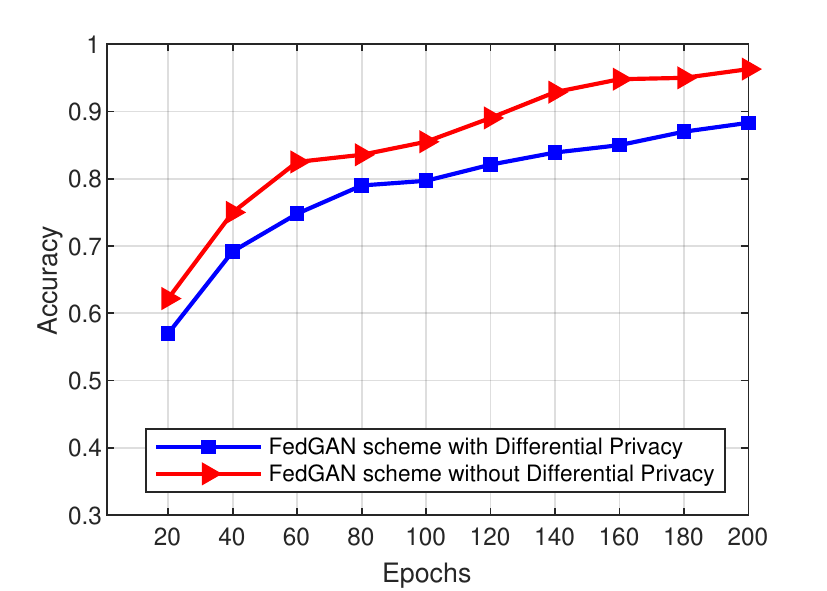} 
		\caption{Training on DarkCOVID dataset. }
	\end{subfigure}%
	~
	\begin{subfigure}[t]{0.24\textwidth}
		\centering
		\includegraphics[width=0.99\linewidth]{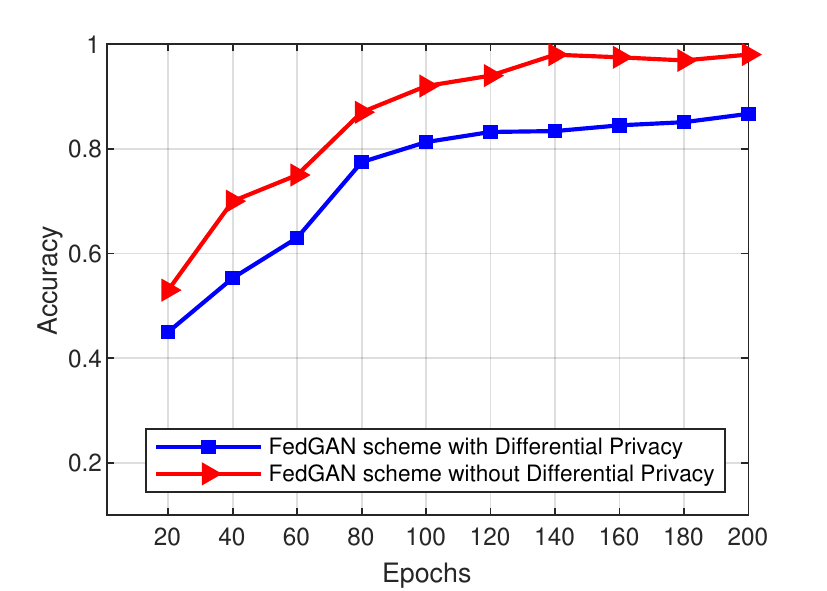} 
		\caption{Training on ChestCOVID dataset.}
	\end{subfigure}
	\caption{{\color{black}Accuracy performance of differential privacy-enabled FedGAN.}}
	\label{Perforcemance_overall-DP}
	\vspace{-0.1in}
\end{figure}
\begin{figure}
	\centering
	\includegraphics [width=0.9\linewidth]{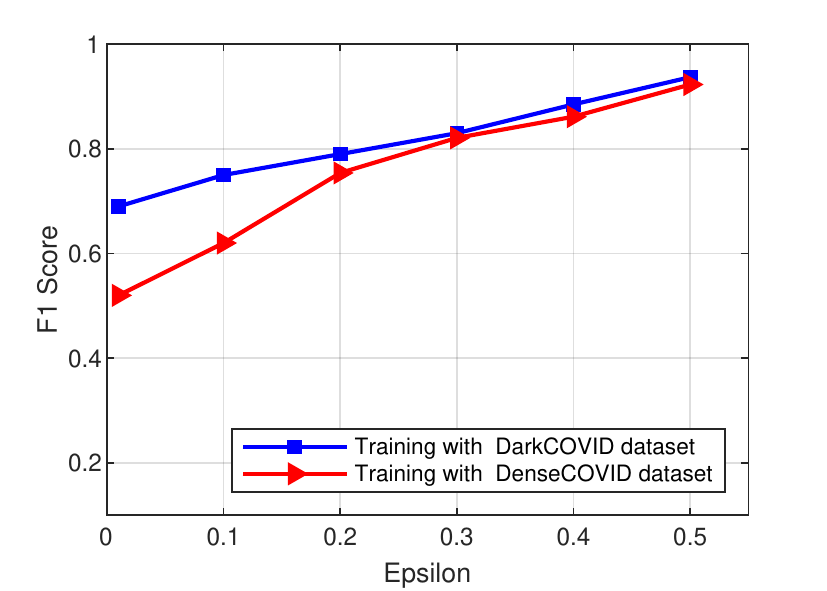}
	\caption{{\color{black}Impacts of $\epsilon$-differential privacy on FedGAN training.} }
	\label{Perforcemance_overall-DP_Epsilon}
		\vspace{-0.1in}
\end{figure}
\begin{figure}
	\centering
	\includegraphics [width=0.95\linewidth]{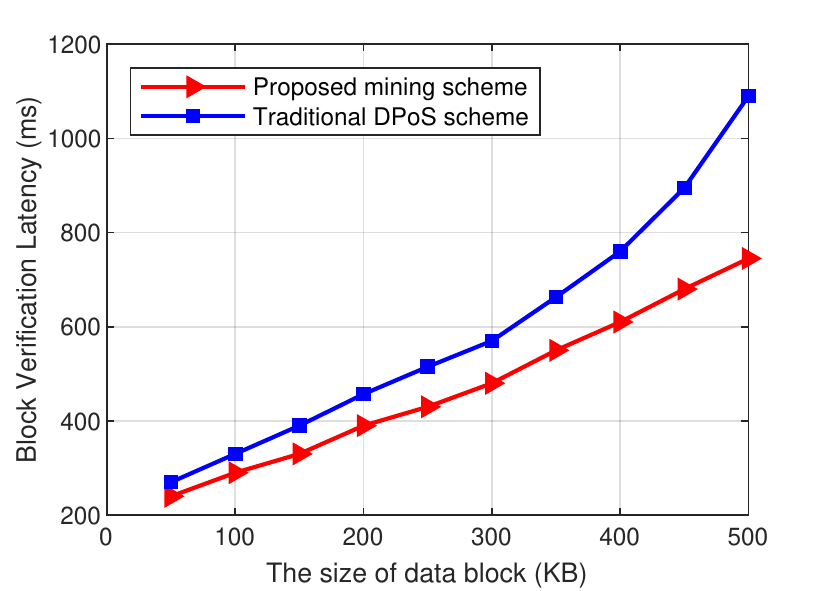}
	\caption{Comparison of block verification latency.}
	\label{Fig:ConsensusLatency_Block}
		\vspace{-0.1in}
\end{figure}

\begin{figure*}[t!]
	\centering
	\begin{subfigure}[t]{0.48\textwidth}
		\centering
		\includegraphics[width=0.99\linewidth]{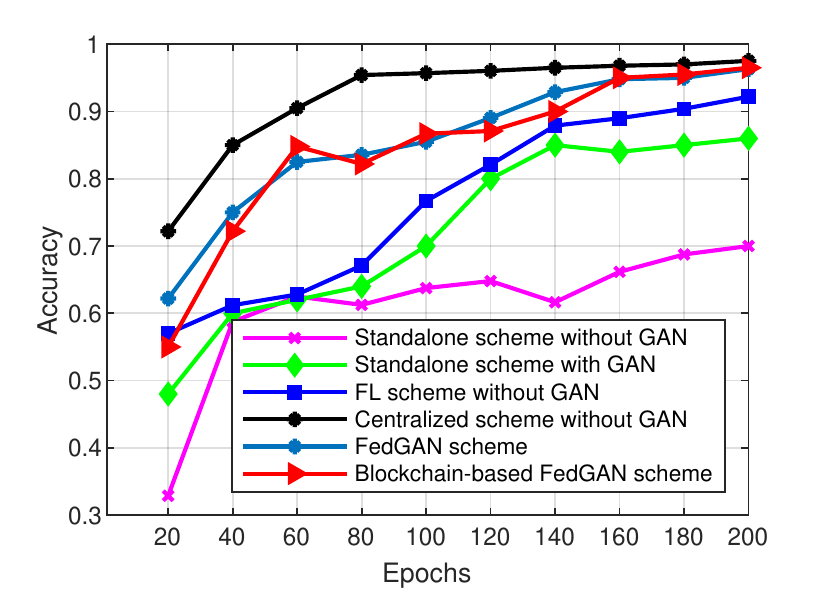} 
		\caption{Accuracy performance on training DarkCOVID dataset. }
	\end{subfigure}%
	~
	\begin{subfigure}[t]{0.48\textwidth}
		\centering
		\includegraphics[width=0.99\linewidth]{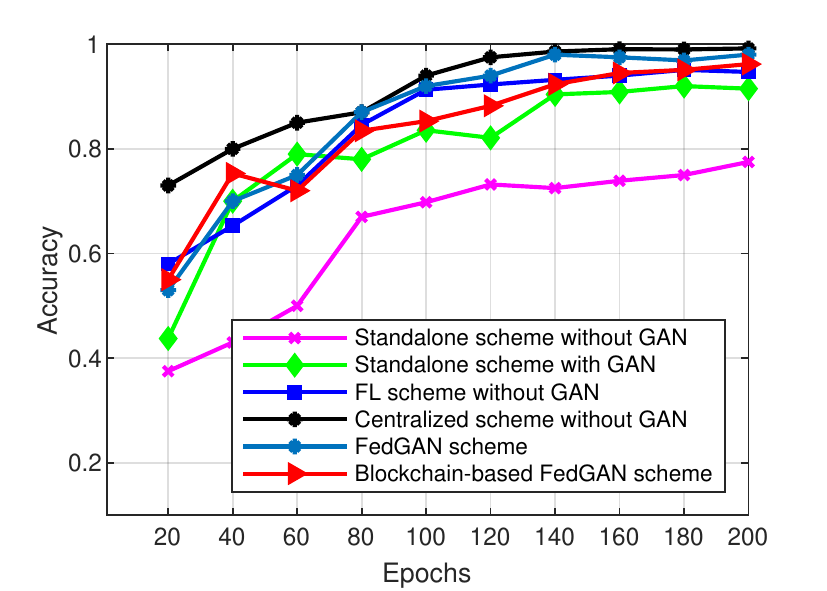} 
		\caption{Accuracy performance on training ChestCOVID dataset. }
	\end{subfigure}
	\caption{\textcolor{black}{Comparison of accuracy performances.}}
	\label{Blockchain_DenseCOVIDset}
	\vspace{-0.1in}
\end{figure*}
We investigate the accuracy performance of the FedGAN scheme with differential privacy, where $\epsilon$ is set to 0.3. As can be seen in Fig.~\ref{Perforcemance_overall-DP}, although differential privacy can provide a degree of privacy to COVID-19 data training, its scheme suffers from a degradation of accuracy performances in both datasets. How to achieve a balance between privacy preservation and data utility (e.g., training accuracy) is still an open problem for further investigation.  

Next, we evaluate the FedGAN scheme in terms of  the data utility performance measured by F1 score when varying the  privacy parameter  $\epsilon$ from 0.01 to 0.5. As shown in Fig.~\ref{Perforcemance_overall-DP_Epsilon}, when $\epsilon$ increases, the level of privacy decreases, and thus enhancing the data utility. This trend is consistent for both DarkCOVID and ChestCOVID datasets. The simulation result also implies that the selection of privacy parameter $\epsilon$ in differential privacy settings plays a significant role in the quality of FL training. 

\begin{figure}
	\centering
	\includegraphics [width=0.99\linewidth]{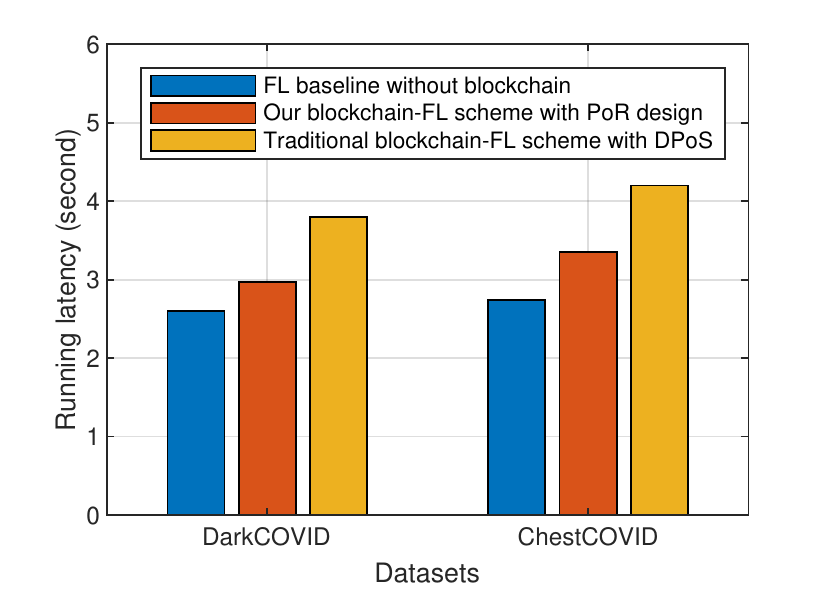}
	\caption{{\color{black} Comparison of running latency.} }
	\label{BarChart_CommunicationLatency}
\end{figure}

\subsection{Performance Evaluations on Blockchain-based FedGAN}

Here, we evaluate the performance of our proposed PoR consensus scheme via numerical simulations  and compare it with the traditional DPoS scheme via the verification block latency metric. We set up 10 transactions per block and vary the numbers of miners from 2 to 100. Motivated by \cite{consensus2}, the mining parameters are set up as follows: edge computation resources $c_m = [10^3-10^6]$ CPU cycles/s, input/output block data sizes $B=500$ KB, $B^{re}=50$ KB, the uplink transmission rate $r_m^u =[100-250]$ kbps, the downlink transmission rate $r_m^d =[100-250]$ kbps, $\xi = 0.5$, $\tau_m =1000$ ms.

\subsubsection{Latency Performance of Block Verification}
We evaluate the block mining latency when varying the size of data block from 50 KB to 500 KB in the  blockchain network with 10 miners. As shown in Fig.~\ref {Fig:ConsensusLatency_Block}, our proposed mining scheme yields a lower verification latency than the traditional DPoS scheme due to our lightweight verification strategy. In particular, our mechanism shows its good advantage when the size of data block is large (e.g., $> 400 KB$), while the DPoS scheme requires much time to verify the large blocks. The simulation results also imply the block mining analysis in section~\ref{Subsection:PoRLatency}. 


\subsubsection{Accuracy Performance}
\textcolor{black}{We investigate the accuracy performance of the blockchain-based FedGAN scheme and compare it with other related schemes. From the simulation results in Fig.~\ref{Blockchain_DenseCOVIDset}, we find that the blockchain-based FedGAN scheme can achieve the competitive accuracy rates with the baseline FedGAN scheme, especially when training with the DarkCOVID dataset. This shows that the integration of blockchain with a decentralized model aggregation enabled by the blockchain consensus does not affect the overall training performance, while providing a high degree of security for the federated data training. Compared with other existing schemes, such as the standalone scheme and the FL scheme without GAN, the blockchain-based FedGAN scheme also achieves much better accuracy performances in both datasets.}

\subsubsection{Performance of Total Running Latency}
We then compare the cost of the blockchain-based FedGAN scheme with our PoR design in terms of the total running latency which consists of the training latency and the mining latency at a global training cycle. For fair comparison, we use the pure FL scheme without blockchain as the baseline which only has the training latency. As indicated in Fig.~\ref{BarChart_CommunicationLatency} for both datasets, our proposed blockchain-based FedGAN scheme can achieve a  relatively  competitive latency performance with the FL baseline scheme and save much running time compared with the traditional blockchain-based FedGAN scheme with the DPoS. For example, our scheme can reduce the running time by 23.2\%  and 19.5\% when training the DarkCOVID and ChestCOVID datasets, respectively, compared with the traditional  scheme with  DPoS. These results are enabled by our advanced consensus design that minimizes the mining latency during the aggregation stage, which leads to the reduction in the overall running time.

\textcolor{black}{ To fully realize FL-GAN in practical healthcare networks, several issues and challenges should be considered. For example, how to ensure efficient resource scheduling for edge nodes to run GAN models is a critical issue since the training of large-scale X-ray images along with mining involvement requires a significant amount of  energy and memory resources. Another challenge can be the lack of motivation of edge nodes in joining the FL-GAN process. An edge node may not be willing to devote its resources to train image datasets and run mining if there is no incentives or rewards. Therefore, it is desirable to develop proper incentive mechanisms to encourage edge nodes from hospitals to participate in the FL-GAN process, which in turn ensures the robustness of the federated health data training. }

\section{Conclusion and Future Work}
\label{Section:Conclusion}
This paper has proposed FedGAN, a novel scheme for COVID-19 detection by enabling the joint design of FL and GAN in a distributed medical network with edge cloud computing. The proposed approach has taken the data augmentation using distributed GANs and the federated data training using FL without sharing actual data for COVID-19 detection. {\color{black}To enhance the privacy in federated COVID-19 data training, we have applied a  differential privacy solution at each hospital institution. We have then proposed a new blockchain-based FedGAN framework for secure COVID-19 data analytics, by decentralizing the FL process over the hospital institutions with a novel mining mechanism.} Our theoretical analysis and numerical simulations have showed that our scheme significantly improves the performances of COVID-19 detection, with the high detection accuracy rate and low running time, compared to the state-of-the-art schemes.

\textcolor{black}{ In future work, it is of interest to extend the proposed FL-blockchain model to other healthcare applications. For example, the integrated FL-blockchain framework can be useful for federated human activity analytics, where wearable sensor devices can collaborate to train a shared human motion classification  model. In such cases, blockchain can be used to establish a decentralized network of sensor devices to share the model updates and coordinate the training without relying on a centralized authority. }

\bibliography{Ref}
\bibliographystyle{IEEEtran}

\end{document}